\documentclass{article}
\usepackage{graphicx} 
\graphicspath{{../}{./}}
\usepackage{amsmath}
\usepackage{amsfonts}
\usepackage{float}
\usepackage{authblk}
\usepackage[labelsep=period]{caption}
\usepackage[skip=\baselineskip,indent=0pt]{parskip}
\renewcommand{\figurename}{Fig.}
\usepackage[a4paper,
            bindingoffset=0.2in,
            left=0.75in,
            right=0.75in,
            top=0.75in,
            bottom=0.75in,
            footskip=.25in]{geometry}

\usepackage[
    backend=bibtex,
    style=numeric-comp,
    sorting=none,
    giveninits=true,
    maxbibnames=5,
    minbibnames=1,
    doi=true,
    isbn=false,
    url=false,
    eprint=false
]{biblatex}
\DeclareNameAlias{author}{family-given}
\DeclareFieldFormat{labelnumberwidth}{#1\addperiod}
\DeclareFieldFormat[article,inproceedings,incollection]{title}{#1}
\DeclareFieldFormat[article]{volume}{\mkbibbold{#1}}
\renewbibmacro{in:}{}
\DeclareDelimFormat{finalnamedelim}{\addspace\&\space}
\AtEveryBibitem{%
    \clearfield{abstract}%
    \clearfield{file}%
    \clearfield{keywords}%
    \clearfield{month}%
    \clearfield{note}%
    \clearfield{number}%
    \clearfield{urldate}%
    \clearlist{language}%
}
\begin{document}

\title{Neural networks as decision trees: an analytical solution for learning and neural selectivity}
\author[1]{Hugo Tissot}
\author[2]{Jonas Ranft}
\author[1]{Yves Boubenec}
\renewcommand{\Affilfont}{\footnotesize}
\affil[1]{Laboratoire des Systèmes Perceptifs, Ecole Normale Supérieure, PSL University, CNRS, Paris, France}
\affil[2]{Institut de Biologie de l’Ecole Normale Supérieure (IBENS), Ecole Normale Supérieure, PSL University, CNRS, INSERM, Paris, France}
\renewcommand{\Authands}{, }
\date{September 21, 2026}
\maketitle

\begin{center}
\small
\textbf{Corresponding author:} Hugo Tissot
(\texttt{hugo.tissot@ens.psl.eu}; \texttt{htissot42@gmail.com})\\
\textbf{Contributing authors:} Yves Boubenec (\texttt{yves.boubenec@ens.fr}); Jonas Ranft
(\texttt{jonas.ranft@ens.psl.eu})
\end{center}

\begin{abstract}
Nonlinear neural networks develop structured internal representations, yet how their geometry is 
determined by the tasks being learned remains poorly understood. Here, we develop an analytical framework for
piecewise-linear feedforward and recurrent networks that links learning, activation-region structure, and
neural selectivity. We show that, at any stationary point of gradient-aligned learning, a nonlinear network
decomposes into local linear regressions over its activation regions. Deviations from the corresponding
least-squares solutions are jointly constrained by network weights shared across regions and vanish in the low-error regime,
yielding an approximate piecewise least-squares decomposition of the task. This structure admits a
decision-tree interpretation, which we recover numerically by fitting trees to predict network activation patterns from the input. 
We further derive how the regions in which neurons are active determine the task statistics they capture, 
thereby organizing neural selectivity into distinct subpopulations. The resulting predictions closely match the
selectivity geometry observed in simulated networks and two empirical neural datasets. Finally, we show that
neural baseline regulates activation-pattern diversity, placing networks along a continuum between coarse,
generalizing representations and fine-grained, expressive representations. Together, these results establish
activation regions as a unifying framework for describing how nonlinear networks decompose task structure into
local computations and how those computations can be revealed through neural activity.
\end{abstract}

\section{Introduction}
A central question at the interface of machine learning theory and systems neuroscience is how the observable
properties of neural activity, such as tuning curves, mixed selectivity, or the geometry of population
responses, are determined by the function computed by a network.
For linear networks, this question has a clean answer: deep linear networks under gradient-aligned learning
dynamics converge to least-squares solutions whose representational geometry is governed by the input-output
covariance structure of the task \cite{saxe_exact_2014, saxe_mathematical_2019}.
Both biological and artificial neural systems, however, are fundamentally \textit{nonlinear}.
Modern architectures rely extensively on activation gating and recurrent interactions, while cortical
circuits exhibit strongly nonlinear response properties.
Extending the linear theory to this nonlinear setting is the central theoretical challenge addressed in this
work.
\\
\\
Piecewise-linear networks, including ReLU architectures, partition the input space into activation regions on
which the network implements an affine map
\cite{chu_exact_2018,balestriero_spline_2018}.
Each region is defined by the configuration of active and inactive hidden units, and inputs that share an
activation pattern share the same effective linear computation.
Previous work has characterized the combinatorial and geometric properties of these regions, including their
number, expressivity, and partition structure of the input space \cite{montufar_number_2014,
serra_bounding_2018, hanin_complexity_2019, goujon_number_2023}, but what these regions compute, and how
learning shapes that computation, has remained largely unaddressed.
In particular, how \textit{regional} computations combine to organize the \textit{global} geometry of neural
representations is unknown.
\\
\\
Here, we develop an analytical framework whose ultimate goal is to understand representation geometry
resulting from learning in nonlinear neural networks with piecewise-affine activations. 
We first show that, at any stationary point of gradient-aligned learning, nonlinear networks decompose into
local linear regressions over their activation regions. We then characterize how deviations from the
corresponding least-squares solutions are constrained by the network weights shared across regions.
In the low-error regime, these deviations vanish, implying that expert nonlinear networks implement an
approximate piecewise least-squares decomposition of the task.
We derive this result for deep feedforward networks and extend it to recurrent networks at equilibrium.
This framework yields a two-part functional interpretation of nonlinear computation: the routing of inputs to
specific neural ensembles and the implementation of a region-specific linear regression by each ensemble.
This decomposition provides a direct link between the activation structure and the geometry of neural
selectivity in real neural data.
\\
\\
A key prediction of this framework is therefore a closed-form solution for the geometry of neural
selectivity, yielding directly testable relationships between task statistics and neural population
structure.
The theory predicts that activation regions induce a modular organization of selectivity across neural
subpopulations: neurons that share activation patterns develop similar selectivity because they implement
the same local input-output mappings, whereas neurons associated with distinct regions specialize toward
different task statistics.
Such modular and task-dependent organization is a recurring feature of biological and artificial neural
representations, where neurons often segregate into functionally specialized subpopulations with distinct
selectivity geometries \cite{rigotti_importance_2013,mante_context-dependent_2013,yang_task_2019,reinert_mouse_2021,dubreuil_role_2022,barbosa_early_2023,tissot_sensorimotor_2026}.
Our framework offers a mechanistic account for the emergence of this structure, with explicit analytical
expressions linking the selectivity covariance of each activation-defined neural subpopulation to task
statistics.
We validate these predictions in trained artificial neural networks and two independent mouse prefrontal and visual cortex
datasets \cite{reinert_mouse_2021,hajnal_shifts_2024}, finding close quantitative agreement between predicted and observed selectivity
geometries.

\section{Deep feedforward network}
\subsection{Setup}

\begin{figure}[H]
\centering
\includegraphics[width=0.5\linewidth]{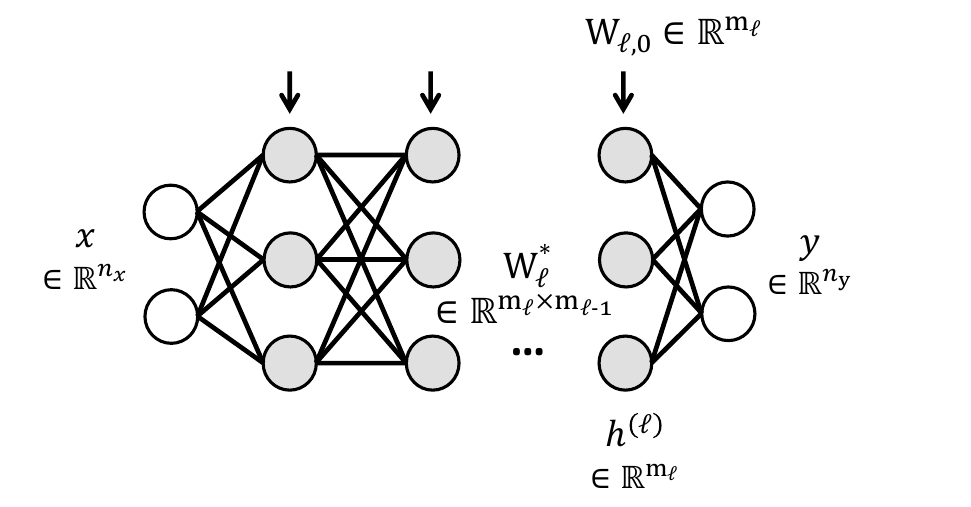}
\caption{Schematic of the feedforward model analyzed in this section}
\end{figure}

We consider an $L$-layer network with input dimension $n_x$, hidden widths $m_1,\ldots,m_{L-1}$, and output
dimension $n_y$.
For compactness, set $m_0:=n_x$ and $m_L:=n_y$.
The network receives a dataset of $p$ input--target pairs
$\{(x_i^*,y_i)\}_{i=1}^p$, where $x_i^*\in\mathbb R^{n_x}$ is the ordinary input of sample $i$ and
$y_i\in\mathbb R^{n_y}$ is its target. The index $i$ labels samples throughout the derivation.
Because the activation functions are piecewise affine, each input selects one affine piece at every hidden
layer.
Let $f_i^{(\ell)}$ denote the piece selected by input $i$ at layer $\ell$.
The corresponding activation region is
\[
S_r
=
\left\{
i\;\middle|\;
f_i^{(\ell)}=f_r^{(\ell)}
\text{ for }1\leq\ell\leq L-1
\right\}.
\]
Thus, inputs belong to the same region when they select the same affine piece at every layer.

Within region $r$, the slopes and intercepts of these pieces are fixed.
Since the activation function is applied independently to each unit, the slopes of the selected affine pieces
at layer $\ell$ can be collected into a diagonal matrix $D_r^{(\ell)}\in\mathbb R^{m_\ell\times m_\ell}$. The
activation function for layer $\ell$ can then be written explicitly within region $r$ as the affine
transformation
\[
\boxed{
f_r^{(\ell)}(z)
=
D_r^{(\ell)}z+f_r^{(\ell)}(0),
\qquad 1\leq\ell\leq L-1.
}
\]
Here $f_r^{(\ell)}(0)$ denotes the intercept of the selected affine piece.
For ReLU, $D_r^{(\ell)}$ is a diagonal binary mask and $f_r^{(\ell)}(0)=0$.
The network is therefore a composition of fixed affine maps within each activation region.

To incorporate the trainable biases into the same matrix expressions as the linear weights, we index the
constant coordinate by $0$ and write
\[
W_\ell
=
\left(
\begin{array}{c|ccc}
w_{10}^{(\ell)}
& w_{11}^{(\ell)} & \cdots & w_{1m_{\ell-1}}^{(\ell)}\\
\vdots & \vdots & \ddots & \vdots\\
w_{m_\ell0}^{(\ell)}
& w_{m_\ell1}^{(\ell)} & \cdots
& w_{m_\ell m_{\ell-1}}^{(\ell)}
\end{array}
\right)
\in\mathbb R^{m_\ell\times(m_{\ell-1}+1)}.
\]
We use the compact block notation
\[
W_\ell
=
\left(
\begin{array}{c|c}
W_{\ell,0}&W_\ell^*
\end{array}
\right),
\]
where $W_{\ell,0}\in\mathbb R^{m_\ell}$ is the trainable bias vector and
$W_\ell^*\in\mathbb R^{m_\ell\times m_{\ell-1}}$ contains the weights connecting layer $\ell-1$ to layer
$\ell$.
The augmented input and activity vectors are ordered with their constant coordinate first.
\[
x_i
=
\begin{pmatrix}
1\\
x_i^*
\end{pmatrix}
\in\mathbb R^{n_x+1},
\qquad
h_i^{(\ell)}
=
\begin{pmatrix}
1\\
h_{i,1}^{(\ell)}\\
\vdots\\
h_{i,m_\ell}^{(\ell)}
\end{pmatrix}
\in\mathbb R^{m_\ell+1}.
\]
\[
h_i^{(0)}=x_i,
\qquad
h_i^{(\ell)}
=
\begin{pmatrix}
1\\
f_i^{(\ell)}\!\left(W_\ell h_i^{(\ell-1)}\right)
\end{pmatrix},
\quad 1\le\ell\le L-1,
\]
The forward pass is
\[
\hat y_i=W_Lh_i^{(L-1)}.
\]
\subsection{Affine form of the network}

Within region $r$, the network is affine.
\begin{equation}
\hat y_i=M_rx_i=A_rx_i^*+c_r,
\qquad
M_r=\left(\begin{array}{c|c}c_r&A_r\end{array}\right).
\label{eq:regional-affine-map}
\end{equation}

\paragraph{Linear part.} Composing the regional slopes and weight matrices gives
\[
\boxed{
A_r
=
W_L^*D_r^{(L-1)}W_{L-1}^*
\cdots
D_r^{(1)}W_1^*.
}
\]

\paragraph{Offset term.} Let $b_r^{(\ell)}$ denote the regional activity offset at layer $\ell$.
It is the activity produced by the regional affine map when the network input is zero.
Starting from
\[
b_r^{(0)}=0,
\]
these offsets obey the recursion
\[
b_r^{(\ell)}
=
f_r^{(\ell)}\!\left(
W_\ell^*b_r^{(\ell-1)}+W_{\ell,0}
\right),
\qquad
1\leq\ell\leq L-1.
\]
The regional output intercept is therefore
\[
\boxed{
c_r
=
W_L^*b_r^{(L-1)}+W_{L,0}.
}
\]
The recursion makes explicit how constant terms introduced at early layers are transformed by every
subsequent layer.

The pair \((A_r,c_r)\) fully characterizes the affine map in region \(r\): 
\(A_r\) is its linear transformation, whereas \(c_r\) is its output offset.

\subsection{Gradient with respect to $W_\ell$}

We now compute the output differential induced by a perturbation \(dW_\ell\) of the weight matrix at layer \(\ell\), within a fixed input region \(r\).
Define the downstream propagation operator
\begin{equation}
\Phi_r^{(\ell\to L)}
:=
\begin{cases}
W_L^*D_r^{(L-1)}W_{L-1}^*
\cdots
D_r^{(\ell+1)}W_{\ell+1}^*D_r^{(\ell)},
&1\le\ell\le L-1,\\
I_{n_y},&\ell=L.
\end{cases}
\label{eq:ff-downstream-operator}
\end{equation}
The chain rule gives the output differential
\begin{equation}
d\hat y_i
=
\Phi_r^{(\ell\to L)}
(dW_\ell)
h_i^{(\ell-1)}.
\label{eq:ff-output-differential}
\end{equation}
This is the standard backpropagation differential.
The operator $\Phi_r^{(\ell\to L)}$ propagates the perturbation from layer $\ell$ to the output, while
$h_i^{(\ell-1)}$ is the presynaptic activity presented to $W_\ell$.
Define the per-sample error $e_i=y_i-M_rx_i$.
The differential of the mean-squared loss $\mathcal L=\frac1p\sum_i\|e_i\|^2$ is
\[
d\mathcal L
=
-\frac2p\sum_i e_i^T d\hat y_i.
\]
Substituting Eq.~\eqref{eq:ff-output-differential} and using the trace identity to isolate \(dW_\ell\) yields
\begin{equation}
\boxed{
\frac{\partial\mathcal L}{\partial W_\ell}
=
-\frac2p
\sum_{r=1}^R\sum_{i\in S_r}
(\Phi_r^{(\ell\to L)})^T
e_i
(h_i^{(\ell-1)})^T.
}
\label{eq:ff-weight-gradient}
\end{equation}
The gradient therefore decomposes into region-wise contributions. Each contribution is the sum across samples 
in that region of outer products between the upstream error signal and the activity entering the layer.
Its first column corresponds to the gradient with respect to the trainable bias $W_{\ell,0}$, while its remaining
columns correspond to the gradient with respect to the weight matrix $W_\ell^*$.

\subsection{Stationarity condition}

For $S_r=\{i_1,\ldots,i_{p_r}\}$, define
\begin{align*}
X_r&=(x_i)_{i\in S_r}
\in\mathbb R^{(n_x+1)\times p_r},\\
Y_r&=(y_i)_{i\in S_r}
\in\mathbb R^{n_y\times p_r},\\
H_r^{(\ell-1)}&=(h_i^{(\ell-1)})_{i\in S_r}
\in\mathbb R^{(m_{\ell-1}+1)\times p_r}.
\end{align*}
Define the augmented upstream propagation operator
\begin{equation}
\Phi_r^{(1\to\ell-1)}
:=
\begin{cases}
\left(
\begin{array}{c|c}
1&0\\ \hline
b_r^{(\ell-1)}
&
D_r^{(\ell-1)}W_{\ell-1}^*
\cdots
D_r^{(1)}W_1^*
\end{array}
\right),
&2\leq\ell\leq L,\\[1ex]
I,&\ell=1,
\end{cases}
\label{eq:ff-upstream-operator}
\end{equation}
Then
\begin{equation}
H_r^{(\ell-1)}=\Phi_r^{(1\to\ell-1)}X_r,
\qquad
E_r=Y_r-M_rX_r.
\label{eq:ff-regional-matrices}
\end{equation}
Stationarity with respect to $W_\ell$ gives
\[
\sum_{r=1}^R
(\Phi_r^{(\ell\to L)})^T
E_r
(H_r^{(\ell-1)})^T
=0.
\]
Using Eq.~\eqref{eq:ff-regional-matrices}, this becomes
\begin{equation}
\sum_{r=1}^R
\underbrace{(\Phi_r^{(\ell\to L)})^T}_{\text{backward pass}}
E_r
X_r^T
(\underbrace{\Phi_r^{(1\to\ell-1)}}_{\text{forward pass}})^T
=0.
\label{eq:ff-stationarity}
\end{equation}
Each term in the sum is the contribution of region \(r\) to the gradient of \(W_\ell\). 
The central factor \(E_rX_r^T\) measures the correlation between prediction errors and inputs within that region. 
The transposed downstream operator on the left propagates these errors backward from the output to layer \(\ell\),
while the upstream operator appearing transposed on the right maps the inputs to the activities entering that layer.
Because \(W_\ell\) is shared across regions, stationarity requires only that these regional gradient contributions cancel across the sum, not that each contribution vanish independently.

\subsection{Least-squares decomposition}

Assuming $X_r$ has full row rank, define the regional affine least-squares estimator
\begin{equation}
\boxed{
M_r^{LS}
:=
Y_rX_r^T
\Sigma_{X_r}^{-1},
\qquad
\Sigma_{X_r}:=X_rX_r^T.
}
\label{eq:regional-ls-estimator}
\end{equation}
Since
\begin{equation}
E_rX_r^T
=
(M_r^{LS}-M_r)\Sigma_{X_r},
\label{eq:regional-ls-identity}
\end{equation}
substitution into Eq.~\eqref{eq:ff-stationarity} gives
\[
\sum_{r=1}^R
(\Phi_r^{(\ell\to L)})^T
(M_r^{LS}-M_r)
\Sigma_{X_r}
(\Phi_r^{(1\to\ell-1)})^T
=0.
\]
Define the regional deviation as the difference between the transformation implemented by the network and
the local least-squares solution,
\[
\boxed{
\Delta_r:=M_r-M_r^{LS},
\qquad
M_r=M_r^{LS}+\Delta_r.
}
\]
Then
\begin{equation}
\boxed{
\sum_{r=1}^R
(\Phi_r^{(\ell\to L)})^T
\Delta_r
\Sigma_{X_r}
(\Phi_r^{(1\to\ell-1)})^T
=0.
}
\label{eq:ff-deviation-balance}
\end{equation}
Thus, $\Delta_r$ measures how far the affine transformation implemented in region $r$ lies from its local
least-squares solution. Because the same matrices $W_\ell$ are shared across regional transformations, these
deviations are not independent, and their weighted contributions must cancel collectively.
In this sense, a nonlinear network at stationarity globally balances regional errors, with $\Delta_r$
quantifying the effect of shared weights on regional least-squares optimality.
In the single-region linear setting, there is no coupling between regions, and stationarity recovers the
least-squares solution, $\Delta=0$ \cite{saxe_exact_2014,saxe_mathematical_2019}.
We next investigate when the deviations $\Delta_r$ also become negligible in nonlinear networks.

\subsection*{Notations}
\begin{center}
\small
\renewcommand{\arraystretch}{1.25}
\begin{tabular}{p{0.23\linewidth}p{0.67\linewidth}}
\hline
Symbol & Meaning \\
\hline
$S_r$, $p_r$ & Activation region $r$ and its number of observed samples. \\
$X_r$, $Y_r$ & Regional augmented input matrix and target matrix; the first row of $X_r$ is constant. \\
$H_r^{(\ell)}$ & Augmented regional activity at layer $\ell$. \\
$M_r=(c_r\mid A_r)$ & Regional affine input-output map, with intercept $c_r$ and linear component $A_r$. \\
$M_r^{LS}$ & Affine least-squares estimator fitted independently within region $r$. \\
$\Delta_r=M_r-M_r^{LS}$ & Deviation of the network map from the regional least-squares estimator. \\
$E_r=Y_r-M_rX_r$ & Regional output-error matrix. \\
$\Sigma_{X_r}=X_rX_r^T$ & Regional augmented input second-moment matrix. \\
$\Phi_r^{(L)}$, $\Phi_r^{(R)}$ & Regional left and right propagation operators; feedforward variants carry
explicit layer ranges. \\
$\ell_r$ & Mean squared prediction error within region $r$. \\
$\beta$, $\beta_S$ & Full and subpopulation-restricted neural selectivity matrices. \\
$\Pi_S$ & Projector onto the neural subpopulation associated with regions $S$. \\
$\tilde X$, $\tilde Y$, $\tilde H$ & Input, target, and neural activity matrices centered across the complete
dataset. \\
\hline
\end{tabular}
\end{center}
\vspace{3mm}

\section{Convergence toward regional least-squares solutions}
We now consider expert networks that achieve low prediction error within each activation region. 
As established above, shared network parameters can shift each regional affine transformation away from its local least-squares solution. 
We now demonstrate that this deviation vanishes as the regional prediction error decreases, so that each regional predictor converges to its corresponding least-squares solution.
Assuming, as above, that $X_r$ has full row rank, define the error as
$$E_r := Y_r - M_r X_r.$$
Then
$$M_r^{LS} = Y_r X_r^T \Sigma_{X_r}^{-1} = (M_r X_r + E_r) X_r^T \Sigma_{X_r}^{-1} = M_r + E_r X_r^T
\Sigma_{X_r}^{-1}.$$
Hence
\begin{equation}
\boxed{\Delta_r = M_r - M_r^{LS} = -E_r X_r^T \Sigma_{X_r}^{-1}.}
\label{eq:regional-deviation-error}
\end{equation}
\\
\\
Let $p_r = |S_r|$ denote the number of samples in region $r$.
Let $\Sigma_X^{(r)}:=\mathbb{E}[xx^T\mid x\in r]$ denote the population second-moment matrix of the
augmented input distribution conditional on region $r$. This matrix captures the regional input statistics
independently of any particular finite sample realization. For i.i.d. augmented inputs within region $r$,
the law of large numbers gives
$$\frac{1}{p_r} X_r X_r^T \xrightarrow{p_r \to \infty} \Sigma_X^{(r)}.$$
Since $\Sigma_X^{(r)}$ is symmetric positive definite, define its smallest eigenvalue as
$\lambda_{\min}^X:=\lambda_{\min}(\Sigma_X^{(r)})>0$. We then obtain
$$\lambda_{\min}(\Sigma_{X_r}) \sim p_r \cdot \lambda_{\min}^X.$$
Because $\Sigma_{X_r}=X_rX_r^T$, the singular-value decomposition of $X_r$ gives
$$\|X_r^T\Sigma_{X_r}^{-1}\|
=\frac{1}{\sqrt{\lambda_{\min}(\Sigma_{X_r})}}
\sim\frac{1}{\sqrt{p_r\lambda_{\min}^X}}.$$
Where $\|\cdot\|$ denotes the spectral norm, Eq.~\eqref{eq:regional-deviation-error} therefore gives
$$\|\Delta_r\| \leq \|E_r\|\,\|X_r^T\Sigma_{X_r}^{-1}\|.$$
We bound the spectral norm of $E_r$ by its Frobenius norm and define the average per-sample loss in region
$r$ as
$$\ell_r := \frac{1}{p_r}\sum_{i \in S_r}\|e_i\|^2 = \frac{1}{p_r}\|E_r\|_F^2$$
so that
$$\|E_r\| \leq \|E_r\|_F = \sqrt{p_r \ell_r}$$
Combining these bounds gives
\begin{equation}
\boxed{
\|\Delta_r\|
\leq
\sqrt{\frac{\ell_r}{\lambda_{\min}^X}}
+o\!\left(\sqrt{\ell_r}\right)
}
\qquad (p_r \to \infty).
\label{eq:regional-low-error-bound}
\end{equation}
The dependence on $\lambda_{\min}^X$ expresses the role of regional input conditioning: the same prediction
error permits a larger deviation from least squares when the samples poorly constrain some input directions
inside a region.
By Eq.~\eqref{eq:regional-low-error-bound}, the deviation from the affine least-squares solution vanishes
whenever the per-region training loss vanishes.
$$p_r \to \infty \quad \text{and} \quad \ell_r \to 0 \implies \|\Delta_r\| \to 0$$
with rate $O(\sqrt{\ell_r})$.
\\
\\
In other words, low training loss within a region forces the network to behave like the optimal affine
transformation on that region's data. This result requires variation along every input direction within the region. If $X_r$ is
rank deficient, convergence can be established only on the subspace spanned by the regional inputs.

\section{Tree interpretation of nonlinear networks and empirical validation of results}

\begin{figure}[H]
\centering
\includegraphics[width=1\linewidth]{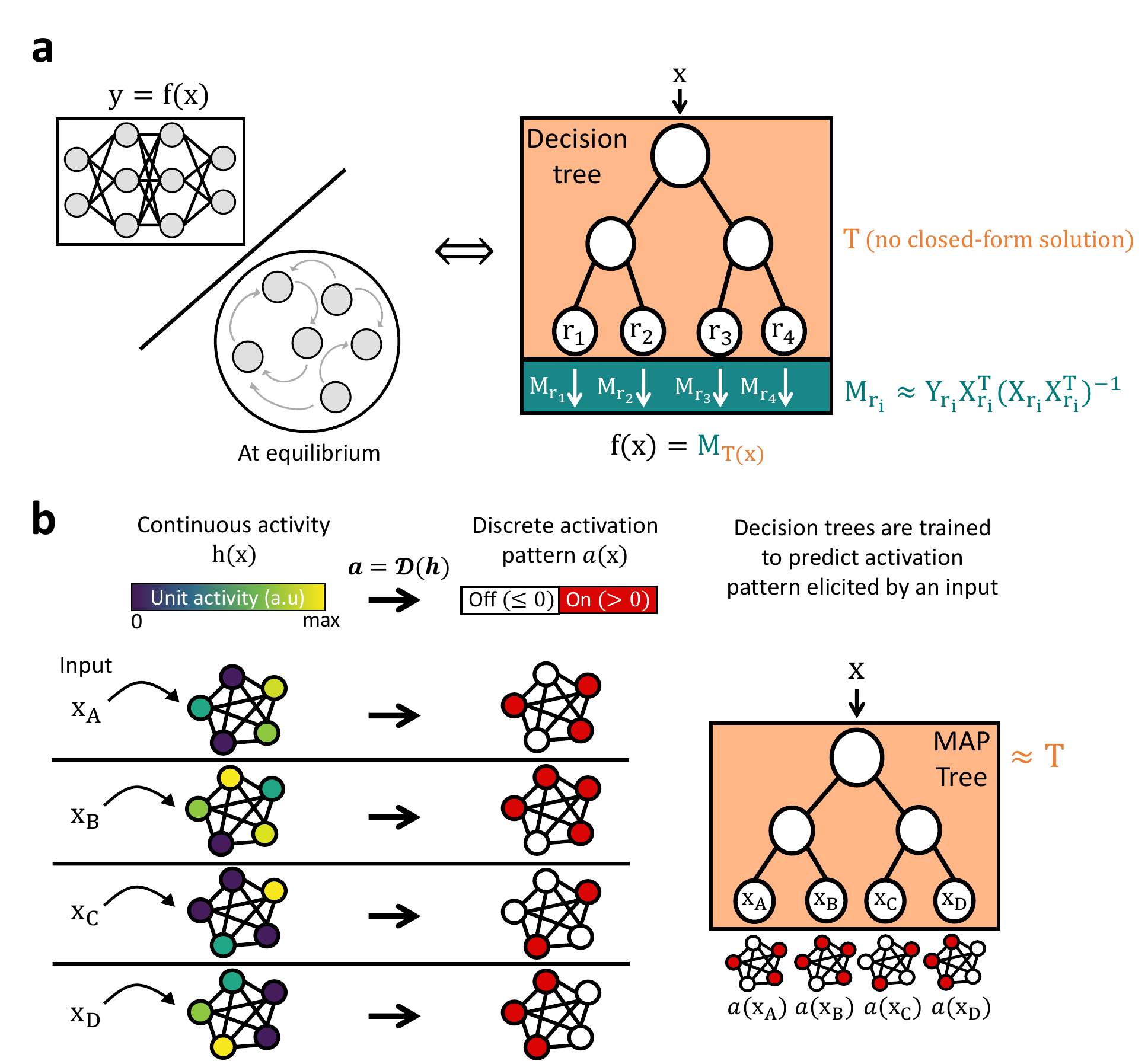}
\caption{\textbf{Tree representation of piecewise-linear neural computations.} 
\textbf{a,} A piecewise-linear feedforward network, or a recurrent network evaluated at equilibrium (see
below), can be decomposed into a routing function $T(x)$ that assigns each input $x$ to an activation region
$r$, followed by the corresponding regional affine transformation $M_r$. 
Although the routing function has no general closed-form expression, the regional transformation approaches
the local least-squares solution $Y_rX_r^T(X_rX_r^T)^{-1}$ in the low-error regime. 
\textbf{b,} Numerical approximation of the routing function. A discretization function \(\mathcal D\) converts continuous activity \(h(x)\) 
into a binary pattern \(a(x)\) indicating whether the activation function of each unit has a zero or nonzero local derivative. For the ReLU example shown here, these states correspond to inactive (\(h\leq0\)) and active (\(h>0\)) units, respectively.
A Main Activation Pattern (MAP) tree is then trained to predict the activation pattern $a(x)$ from the input $x$,
providing an explicit approximation of $T(x)$ and assigning inputs to activation-defined regions.}
\label{fig:tree_equivalence}
\end{figure}

The previous results imply that piecewise-linear networks admit a simple functional interpretation.
Since the network implements an affine transformation within each activation region, its global computation
can be functionally decomposed into two stages: first, the input is routed toward an activation region through a
hierarchical partition of the input space; second, a region-specific affine transformation is applied to produce the output.
This interpretation is illustrated in Fig.~\ref{fig:tree_equivalence}a.
Under this view, a piecewise-linear network is functionally equivalent to a decision tree whose leaves
correspond to activation regions \(r\), each associated with a local affine transformation \(M_r\). 
Related interpretations of piecewise-linear feedforward networks as decision trees have been described
previously \cite{nguyen_towards_2021,aytekin_neural_2022}.
Our analysis further shows that learning drives these regional transformations toward the corresponding
affine least-squares solutions,
\[
M_r
\approx
Y_rX_r^\top(X_rX_r^\top)^{-1},
\]
so that each leaf of the tree behaves as a locally optimal affine transformation specialized to the
input-output statistics of its region.

Importantly, the derivations above characterize the computation performed \emph{within} each activation
region, but do not provide a closed-form description of how these regions are distributed across the input
space. To make practical use of the theoretical predictions, the underlying routing process must therefore be
approximated numerically. Because each activation region is defined by its activation pattern, a natural approximation of the latent
routing structure is obtained by explicitly predicting these patterns from the input.
We therefore trained decision trees to predict the dominant activation pattern elicited by a stimulus, yielding
what we term \emph{Main Activation Pattern} (MAP) trees (Fig.~\ref{fig:tree_equivalence}b).
For artificial ReLU networks, units are active whenever their rectified activity is positive and inactive
otherwise, so the binary vector directly identifies the corresponding activation region.
For recorded neural activity, which has no known ReLU-like absolute threshold, activity states are instead
defined relative to each neuron's mean firing rate, as detailed in Methods.
The MAP tree approximates the hidden partition structure implemented by the network by learning the mapping
from inputs to activation patterns, thereby providing an explicit approximation of the latent decision
process underlying nonlinear computation.

\begin{figure}[H]
\centering
\includegraphics[width=1\linewidth]{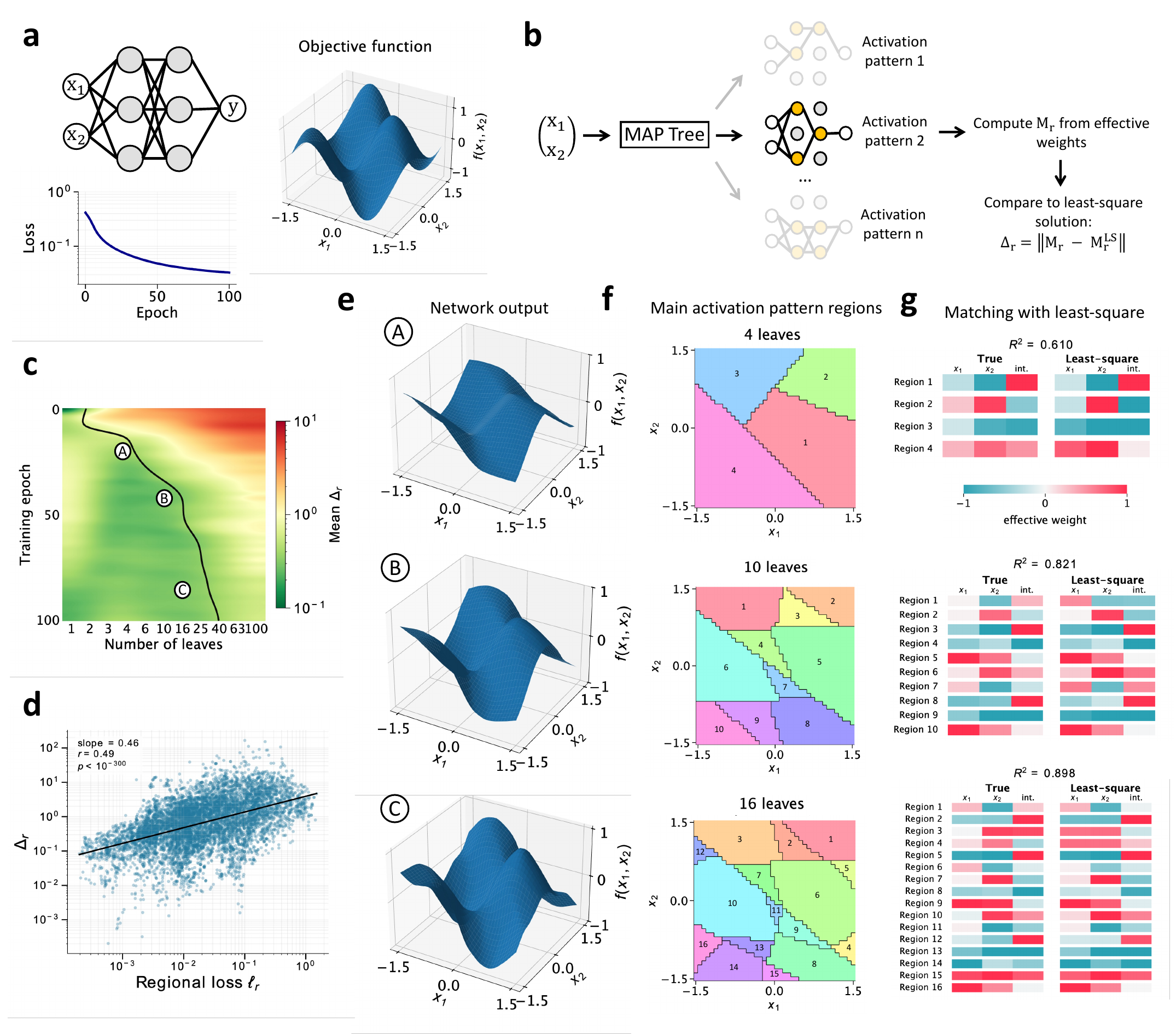}
\caption{\textbf{Convergence of regional transformations toward least-squares solutions in a feedforward
network.} \textbf{a,} Feedforward network trained to approximate a nonlinear two-dimensional target function;
the training loss decreases over epochs. 
\textbf{b,} Analysis procedure. At each training stage, a MAP tree predicts the activation pattern from
$(x_1,x_2)$ and partitions the input space into regions. For each region $r$, the affine transformation $M_r$
is computed from the network's effective weights and compared with the regional least-squares solution
$M_r^{LS}$ through the magnitude of the deviation, $\lVert\Delta_r\rVert=\lVert M_r-M_r^{LS}\rVert$. 
\textbf{c,} Mean regional deviation magnitude $\lVert\Delta_r\rVert$ as a function of training epoch and
number of MAP-tree leaves. The black trajectory marks the critical tree depth beyond which \(\Delta_r\)
increases, illustrating the finest granularity of the least-squares solutions captured by the network during
learning. A-C indicate the stages illustrated in \textbf{e-g}. 
\textbf{d,} Regional deviation magnitude plotted against regional loss across training stages and MAP-tree
regions. The log-log relationship has slope $0.46$ ($r=0.49$, $p<10^{-300}$), close to the square-root
dependence predicted by the bound on $\lVert\Delta_r\rVert$. 
\textbf{e,} Network output surfaces at stages A-C. \textbf{f,} Corresponding MAP-tree partitions of the input
space into 4, 10, and 16 leaves. Colors identify distinct regions, and numbers denote the region indices used
in \textbf{g}. 
\textbf{g,} Regional coefficients of the transformations implemented by the network (True; columns $x_1$,
$x_2$, and intercept) and their least-squares counterparts for the partitions in \textbf{f}. Agreement
increases as training progresses and regional error decreases ($R^2=0.610$, $0.821$, and $0.898$ for A-C,
respectively).}
\label{fig:feedforward_simulation}
\end{figure}

We first tested the regional least-squares prediction in feedforward networks trained to approximate a
nonlinear function of two continuous inputs (Fig.~\ref{fig:feedforward_simulation}a). At successive stages of
learning, we fitted MAP trees with increasing numbers of leaves and compared the affine map implemented in
each inferred region with the corresponding local least-squares solution
(Fig.~\ref{fig:feedforward_simulation}b). This analysis tested both whether regional transformations approach
least squares during learning and how finely this decomposition can be resolved by the network.

Regional deviations decreased over training for a broad range of tree sizes
(Fig.~\ref{fig:feedforward_simulation}c). For each training stage, however, increasing tree depth beyond a
critical value caused the deviation to rise again, identifying the finest partition at which the learned
regional maps remained well described by local least squares. Across training stages and regions, the
deviation scaled with regional loss with a log-log slope of $0.46$, close to the square-root dependence
predicted analytically (Fig.~\ref{fig:feedforward_simulation}d). The same progression was visible directly in
the learned surfaces and regional coefficients: agreement between network and least-squares transformations
increased from $R^2=0.610$ to $0.898$ across the three illustrated stages
(Fig.~\ref{fig:feedforward_simulation}e-g). Thus, learning progressively resolved the nonlinear target into
finer regional regressions.

\section{Recurrent neural network at equilibrium}

We next extend the regional least-squares framework to recurrent neural networks at equilibrium.
A recurrent network admits a fixed-point representation that, upon unfolding, is formally equivalent to a
deep feedforward network with tied affine weights.
This equivalence allows us to carry the least-squares characterization over to the recurrent setting.
The recurrent setting is of particular interest because it provides a natural model of cortical dynamics,
where neurons interact through reciprocal connections and settle into a steady state in response to a
sustained input.
The aim of this section is therefore to yield predictions that are directly testable against
electrophysiological recordings in biological neural networks.
\subsection{Setup}

\begin{figure}[H]
\centering
\includegraphics[width=0.5\linewidth]{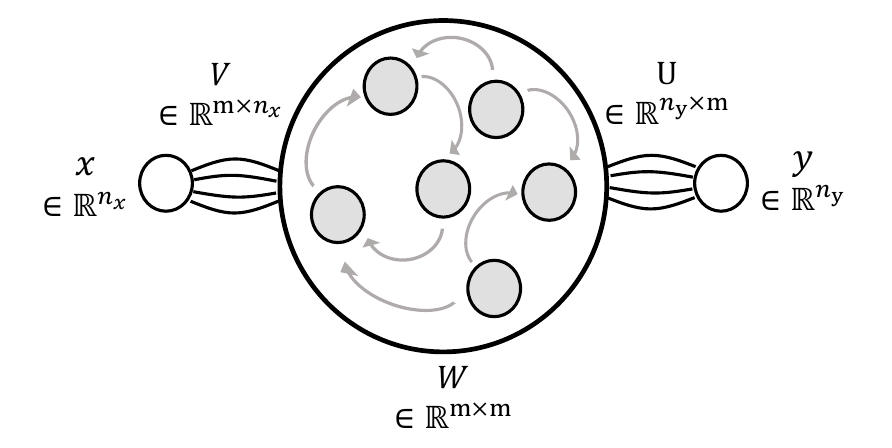}
\caption{Schematic of the recurrent model analyzed in this section}
\end{figure}

For a sustained ordinary input $x_i^*\in\mathbb R^{n_x}$, the network state $\kappa_i(t)\in\mathbb R^m$ evolves according to the
continuous-time dynamics
\[
\tau\frac{\mathrm d\kappa_i}{\mathrm dt}
=
-\kappa_i+W_0+W^*f(\kappa_i)+V^*x_i^*,
\]
where $\tau$ is the neural time constant, $W_0$ is the recurrent bias, $W^*$ is the recurrent weight matrix,
$V^*$ is the input projection, and $f$ is a coordinate-wise piecewise-affine activation function.
We focus on stable equilibria of these dynamics, obtained by setting $\mathrm d\kappa_i/\mathrm dt=0$, and
henceforth let $\kappa_i$ denote the equilibrium state elicited by input $x_i^*$.
As in the feedforward network, the piecewise-affine activation partitions the inputs
according to the affine piece selected at equilibrium.
Writing $f_i$ for the piece selected by input $i$, the recurrent activation regions are
\[
S_r
=
\left\{
i\;\middle|\;f_i=f_r
\right\}.
\]
Within region $r$, the activation has a fixed diagonal slope matrix $D_r\in\mathbb R^{m\times m}$ and a fixed
intercept.
\[
\boxed{
f_r(\kappa)=D_r\kappa+f_r(0).
}
\]
For ReLU, $D_r$ is a diagonal binary mask and $f_r(0)=0$.

We retain the augmented-coordinate convention introduced above and write
\[
x_i
=
\begin{pmatrix}
1\\
x_i^*
\end{pmatrix}
\in\mathbb R^{n_x+1}.
\]
Thus, $x_i^*$ denotes the ordinary input, whereas $x_i$ denotes its augmented representation.
The recurrent and input weight matrices are
\[
W
=
\left(
\begin{array}{c|ccc}
w_{10} & w_{11} & \cdots & w_{1m}\\
\vdots & \vdots & \ddots & \vdots\\
w_{m0} & w_{m1} & \cdots & w_{mm}
\end{array}
\right)
\in\mathbb R^{m\times(m+1)},
\]
and
\[
V
=
\left(
\begin{array}{c|ccc}
0 & v_{11} & \cdots & v_{1n_x}\\
\vdots & \vdots & \ddots & \vdots\\
0 & v_{m1} & \cdots & v_{mn_x}
\end{array}
\right)
\in\mathbb R^{m\times(n_x+1)}.
\]
In compact block notation,
\[
W=(W_0\mid W^*),
\qquad
V=(0\mid V^*),
\]
where $W_0$ is the trainable recurrent bias, $W^*$ is the recurrent weight matrix, and $V^*$ is the
input weight matrix.
The first column of $V$ vanishes because the constant input coordinate is already represented in $W$.
Defining
\[
h_i
=
\begin{pmatrix}
1\\
f_i(\kappa_i)
\end{pmatrix},
\]
the equilibrium equation for the full recurrent network is
\[
\kappa_i=Wh_i+Vx_i.
\]
For $i\in S_r$, this becomes
\[
\kappa_i
=W_0+W^*f_r(\kappa_i)+V^*x_i^*.
\]
Let $b_r$ denote the regional preactivation offset.
It satisfies
\[
b_r=W_0+W^*f_r(b_r).
\]
Define
\[
Z_r:=I-W^*D_r.
\]
Subtracting the regional offset equation from the equilibrium equation gives
\[
Z_r(\kappa_i-b_r)=Vx_i.
\]
If $Z_r$ is invertible, the regional fixed point is therefore
\begin{equation}
\boxed{
\kappa_i=b_r+Z_r^{-1}Vx_i.
}
\label{eq:rnn-regional-fixed-point}
\end{equation}
The inverse $Z_r^{-1}=(I-W^*D_r)^{-1}$ is the recurrent counterpart of a feedforward propagation operator.
It captures the cumulative effect of recurrent interactions on an input perturbation within region \(r\).

\subsection{Gradient and stationarity condition}

For the readout $\hat y_i=Uf_r(\kappa_i)$, with $U\in\mathbb R^{n_y\times m}$, the regional map is affine.
Since
\[
f_r(\kappa_i)
=
f_r(b_r)+D_rZ_r^{-1}Vx_i,
\]
defining
\[
A_r:=UD_rZ_r^{-1}V^*,
\qquad
c_r:=Uf_r(b_r),
\]
the regional readout takes the affine form
\[
\hat y_i
=
A_rx_i^* + c_r
=
M_rx_i,
\qquad
M_r:=\left(\begin{array}{c|c}c_r&A_r\end{array}\right).
\]
Differentiating the fixed-point equation gives $d\kappa_i=Z_r^{-1}(dW)h_i$ and hence
\[
d\hat{y}_i
=
UD_rZ_r^{-1}(dW)h_i.
\]
Defining the sample error $e_i=y_i-M_rx_i$ and the left propagation operator
$\Phi_r^{(L)}:=UD_rZ_r^{-1}$, the gradient becomes
\begin{equation}
\boxed{\frac{\partial \mathcal{L}}{\partial W}
=
-\frac{2}{p}
\sum_{r=1}^R\sum_{i\in S_r}
(\Phi_r^{(L)})^T e_i\, (h_i)^T.}
\label{eq:rnn-weight-gradient}
\end{equation}
To express the activity term $h_i$ in terms of the input, we similarly define the augmented right propagation
operator
\begin{equation}
\Phi_r^{(R)}
:=
\left(
\begin{array}{c|c}
1 & 0\\ \hline
f_r(b_r)
&
D_rZ_r^{-1}V^*
\end{array}
\right).
\label{eq:rnn-right-operator}
\end{equation}
The equilibrium activity therefore satisfies
\[
H_r=\Phi_r^{(R)}X_r.
\]
Setting $\partial\mathcal{L}/\partial W=0$ and using the regional least-squares relation in
Eq.~\eqref{eq:regional-ls-identity} gives the recurrent stationarity condition
\begin{equation}
\sum_{r=1}^R
(\Phi_r^{(L)})^T
(M_r^{LS}-M_r)
\Sigma_{X_r}
(\Phi_r^{(R)})^T
=0.
\label{eq:rnn-stationarity}
\end{equation}
Thus
\[\boxed{M_r = M_r^{LS} + \Delta_r}\]
with deviation $\Delta_r$ meeting the constraint obtained from Eq.~\eqref{eq:rnn-stationarity},
\begin{equation}
\boxed{
\sum_{r=1}^R
(\Phi_r^{(L)})^T
\Delta_r
\Sigma_{X_r}
(\Phi_r^{(R)})^T
=0.
}
\label{eq:rnn-deviation-balance}
\end{equation}

This condition has the same structure as its feedforward counterpart.
The left operator propagates output errors through the recurrent dynamics, whereas the right operator maps
inputs to equilibrium activities.
Global stationarity therefore constrains a weighted sum of regional affine deviations, with the weights now
determined by the recurrent interactions rather than a finite product of layer matrices.

\paragraph{Special case: fixed zero bias.}
If $W_0=0$ and the activation maps zero to zero, then $b_r=0$ and $c_r=0$.
The augmented stationarity condition then reduces to its non-augmented form.

\begin{figure}[H]
\centering
\includegraphics[width=1\linewidth]{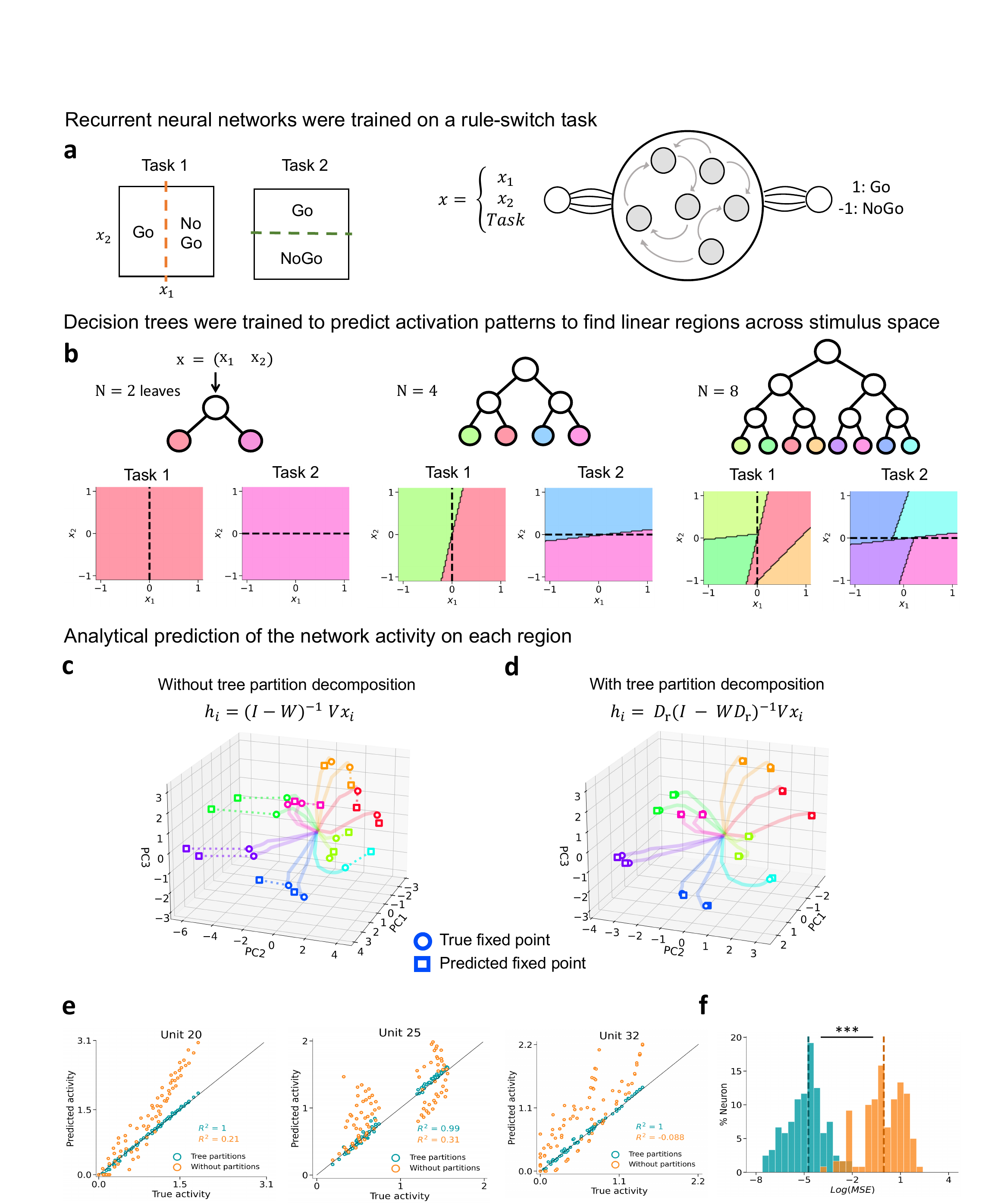}
\caption{\textbf{Activation-region decomposition predicts equilibrium activity in recurrent neural networks.}
\textit{(Caption on next page.)}}
\label{fig:rnn_simulation}
\end{figure}

Because the recurrent map has the same regional least-squares decomposition, the low-error convergence
argument derived above applies without modification.
We tested this extension in recurrent networks trained on a rule-switch task in which a contextual cue made
either $x_1$ or $x_2$ relevant for the Go/NoGo decision (Fig.~\ref{fig:rnn_simulation}a). MAP trees fitted to
equilibrium activation patterns recovered task-dependent partitions of stimulus space whose resolution
increased with the number of leaves (Fig.~\ref{fig:rnn_simulation}b). We then asked whether these partitions
were required to predict the equilibrium reached by the network.
\
\
A single global linear fixed-point approximation systematically displaced the predicted equilibria from the
true network states (Fig.~\ref{fig:rnn_simulation}c). In contrast, using the activation mask associated with
each MAP-tree region brought the analytical fixed points close to the observed equilibria
(Fig.~\ref{fig:rnn_simulation}d). This improvement held at the level of individual units: regional predictions
closely followed the identity line in the examples shown, whereas the global approximation failed to capture
their input-dependent activity (Fig.~\ref{fig:rnn_simulation}e). Across the population, incorporating the
regional partition strongly reduced prediction error (Fig.~\ref{fig:rnn_simulation}f, $W=0$, $p=1.82\times10^{-12}$). These results establish
that the regional decomposition extends from feedforward transformations to the equilibrium activity of
recurrent networks.

\clearpage
\noindent\textbf{Fig. 5. Activation-region decomposition predicts equilibrium activity in recurrent neural
networks.}
\textbf{a,} Recurrent networks were trained on a rule-switch task in which the relevant decision boundary
depends on a task cue: Task 1 separates Go and NoGo responses according to $x_1$, whereas Task 2 separates
them according to $x_2$. 
The network receives $(x_1,x_2,\mathrm{Task})$ and produces readouts for the binary Go/NoGo decision after reaching equilibrium. 
\textbf{b,} MAP trees with 2, 4, or 8 leaves were fitted to the equilibrium activation patterns. Colored
areas show the resulting activation-defined partition of stimulus space separately for the two tasks; dashed
lines show the task decision boundaries. 
\textbf{c,} Equilibrium activity predicted using a single global linear fixed-point approximation, without
decomposing the input space into MAP-tree regions. 
True and predicted fixed points are shown in principal-component space by circles and squares, respectively;
colors identify activation regions and trajectories show the network dynamics. 
\textbf{d,} Same comparison using the region-specific fixed-point solution determined by the activation mask
$D_r$. Accounting for the regional partition brings the predicted fixed points close to the true equilibria. 
\textbf{e,} Predicted versus true equilibrium activity for three example units. Blue points use the MAP-tree
partition and orange points use the global approximation; the diagonal indicates equality and the displayed
$R^2$ values quantify prediction accuracy. 
\textbf{f,} Distribution across neurons of the log mean-squared error between predicted and true equilibrium
activity, with and without the MAP-tree partition. Dashed vertical lines indicate the median of each
distribution. Errors were lower with the MAP-tree partition (two-sided paired Wilcoxon signed-rank test
across 40 neurons, $W=0$, $p=1.82\times10^{-12}$).

\section{How neural selectivities are shaped by learning}

We now apply the low-error result derived above to the geometry of neural selectivity.
We consider a network with ReLU activation functions.
In the large-sample, low-error regime, $\Delta_r\to0$ and each regional affine map approaches its local
affine least-squares solution.
In this section, $X$ and $H$ denote ordinary, non-augmented input and activity matrices.

\subsection{Linear encoding model}

To characterize the geometry of neural selectivity, we consider the linear encoding model
\[
\tilde H=\beta\tilde X+\epsilon.
\]
Here $\tilde X$ and $\tilde H$ are centered across the complete dataset, and $\beta$ contains the linear
selectivity of every neuron to every input dimension.
This is the standard encoding-model description used to quantify stimulus selectivity from neural recordings.
Writing
\[
\Sigma_{\tilde X}:=\tilde X\tilde X^T,
\]
the least-squares estimator and its Gram matrix are
\[
\beta=\tilde H\tilde X^T\Sigma_{\tilde X}^{-1},
\qquad
\beta^T\beta
=
\Sigma_{\tilde X}^{-1}
\tilde X\tilde H^T\tilde H\tilde X^T
\Sigma_{\tilde X}^{-1}.
\]
The matrix $\beta^T\beta$ describes the geometry of these selectivities in input space.
Its diagonal entries measure the total selectivity associated with each input dimension, while its
off-diagonal entries measure the alignment between selectivity directions.
Characterizing this geometry therefore reduces to understanding the activity Gram matrix $\tilde H^T\tilde H$.

\subsection{Block decomposition over activation regions}

Piecewise-linear networks partition the input space into activation regions.
For the recurrent model, write
\[
H=(H_1,\ldots,H_R),
\]
where $H_r=(f_r(\kappa_i))_{i\in S_r}$ is the activated equilibrium activity in region $r$.
Using the fixed-point expression above,
\[
H_r
=
D_rZ_r^{-1}V^*X_r
+
f_r(b_r)\mathbf1^T.
\]
Define the global means
\[
\bar x:=\frac1pX\mathbf1,
\qquad
\bar h:=\frac1pH\mathbf1,
\qquad
p:=\sum_rp_r,
\]
and the centered regional blocks
\[
\tilde X_r:=X_r-\bar x\mathbf1^T,
\qquad
\tilde H_r:=H_r-\bar h\mathbf1^T.
\]
Writing
\[
\tilde X=(\tilde X_1,\ldots,\tilde X_R),
\qquad
\tilde H=(\tilde H_1,\ldots,\tilde H_R),
\]
we obtain
\[
\tilde X\tilde H^T\tilde H\tilde X^T
=
\sum_{r,q}
\tilde X_r\tilde H_r^T
\tilde H_q\tilde X_q^T.
\]
Thus the selectivity geometry decomposes into pairwise interactions between activation regions.
Diagonal terms compare activity patterns within the same region, whereas off-diagonal terms measure how the
representations associated with two different regions align.

\subsection{Selectivity geometry after learning}

For the targets, define
\[
\bar y:=\frac1pY\mathbf1,
\qquad
\tilde Y_r:=Y_r-\bar y\mathbf1^T.
\]
Within region $r$, the readout is the affine map
\[
\widehat Y_r=A_rX_r+c_r\mathbf1^T.
\]
In the large-sample, low-error regime established above, this map approaches
the corresponding affine least-squares solution.  The slope and intercept
conditions give
\[
A_r\bar x_r+c_r\approx\bar y_r,
\]
and
\begin{align*}
\left(\widehat Y_r-\bar y\mathbf1^T\right)\tilde X_r^T
&\approx
\tilde Y_r\tilde X_r^T.
\end{align*}
Since $\widehat Y_r=UH_r$, this yields
\[
UH_r\tilde X_r^T
\approx
\tilde Y_r\tilde X_r^T.
\]

To relate this output-space constraint to neural activity, we assume that the readout matrix
\[
U\in\mathbb R^{n_y\times m}
\]
has normalized orthogonal rows, such that
\[
UU^T=I.
\]
This relation does not uniquely determine $H_r$, since activity components in
$\ker(U)$ are invisible to the output.  However, activity orthogonal to the
readout subspace contributes no functional signal while still incurring
metabolic cost.  Under activity regularization, stationarity therefore favors
minimum-energy representations concentrated in the row space of $U$.  Under
this assumption,
\[
H_r\tilde X_r^T
\approx
U^T\tilde Y_r\tilde X_r^T.
\]
For two regions $r$ and $q$, this representation gives
\[
\tilde X_rH_r^TH_q\tilde X_q^T
\approx
\tilde X_r\tilde Y_r^TU U^T\tilde Y_q\tilde X_q^T.
\]
Using $UU^T=I$, pairwise interactions between regional activity patterns are
therefore determined by the alignment of their output-input covariance
matrices.  Defining
\[
\Sigma_{\tilde Y_r\tilde X_r}:=\tilde Y_r\tilde X_r^T,
\]
we obtain
\[
\tilde X_rH_r^TH_q\tilde X_q^T
\approx
\Sigma_{\tilde Y_r\tilde X_r}^T
\Sigma_{\tilde Y_q\tilde X_q}.
\]
Consequently,
\[
\beta^T\beta
\approx
\Sigma_{\tilde X}^{-1}
\left(
\sum_{r,q}
\Sigma_{\tilde Y_r\tilde X_r}^T
\Sigma_{\tilde Y_q\tilde X_q}
\right)
\Sigma_{\tilde X}^{-1}.
\]
Defining
\[
\Sigma_{\tilde Y\tilde X}
:=\sum_r\Sigma_{\tilde Y_r\tilde X_r},
\]
we obtain
\[
\boxed{
\beta^T\beta
\approx
\Sigma_{\tilde X}^{-1}
\Sigma_{\tilde Y\tilde X}^T\Sigma_{\tilde Y\tilde X}
\Sigma_{\tilde X}^{-1}.
}
\]
The global selectivity geometry is therefore inherited from the total
output-input covariance.  In this sense, a nonlinear network recovers the
same covariance structure as a linear model after its region-specific
solutions are aggregated.

\subsection{Modular decomposition over neural subpopulations}

We now characterize the geometry of selectivity over neural subpopulations.
Let $\Pi_S$ project onto neurons whose activation slope is nonzero only in
regions belonging to $S$, and define $\beta_S:=\Pi_S\beta$.  Within an
activation region, each ReLU unit is either active, with nonzero slope, or
inactive, with zero output.  A neuron selected by $\Pi_S$ is active only in
regions belonging to $S$ and is therefore inactive in every region
$r\notin S$.  Consequently,
\[
\Pi_SH_r=0
\qquad
\text{for }r\notin S.
\]
Since
$\tilde X\mathbf1=0$, centering the activity does not change its covariance
with the input.
\[
\tilde X\tilde H^T
=
\tilde XH^T.
\]
It follows that
\[
\beta_S^T\beta_S
=
\Sigma_{\tilde X}^{-1}
\left(
\sum_{r,q\in S}
\tilde X_rH_r^T\Pi_SH_q\tilde X_q^T
\right)
\Sigma_{\tilde X}^{-1}.
\]
Thus neurons selected by $\Pi_S$ contribute no activity outside the regions
on which their activation slope is nonzero.  The global sum over regions
therefore reduces to interactions entirely within $S$.  Using the
minimum-energy representation gives
\[
\tilde X_rH_r^T\Pi_SH_q\tilde X_q^T
\approx
\Sigma_{\tilde Y_r\tilde X_r}^T
U\Pi_SU^T
\Sigma_{\tilde Y_q\tilde X_q}.
\]
Assuming that the activation-defined subpopulation is sufficiently large and
isotropic in readout space, so that
\[
U\Pi_SU^T\propto I.
\]
Then
\[
\beta_S^T\beta_S
\propto
\Sigma_{\tilde X}^{-1}
\left(
\sum_{r,q\in S}
\Sigma_{\tilde Y_r\tilde X_r}^T
\Sigma_{\tilde Y_q\tilde X_q}
\right)
\Sigma_{\tilde X}^{-1}.
\]
Defining
\[
\Sigma_{\tilde Y_S\tilde X_S}
:=\sum_{r\in S}\Sigma_{\tilde Y_r\tilde X_r},
\]
gives
\[
\boxed{
\beta_S^T\beta_S
\propto
\Sigma_{\tilde X}^{-1}
\Sigma_{\tilde Y_S\tilde X_S}^T
\Sigma_{\tilde Y_S\tilde X_S}
\Sigma_{\tilde X}^{-1}.
}
\]
This assigns a distinct covariance geometry to each activation-defined neural
subpopulation.  Each subpopulation behaves as a local predictor specialized
to the input-output statistics of the regions in $S$, while the outer
factors $\Sigma_{\tilde X}^{-1}$ account for correlations in the complete stimulus
distribution.  Under input whitening, $\Sigma_{\tilde X}=I$, this reduces to
$\beta_S^T\beta_S\propto
\Sigma_{\tilde Y_S\tilde X_S}^T
\Sigma_{\tilde Y_S\tilde X_S}$.

\subsection{Validation of selectivity predictions across simulated and empirical networks}

\begin{figure}[H]
\centering
\includegraphics[width=1\linewidth]{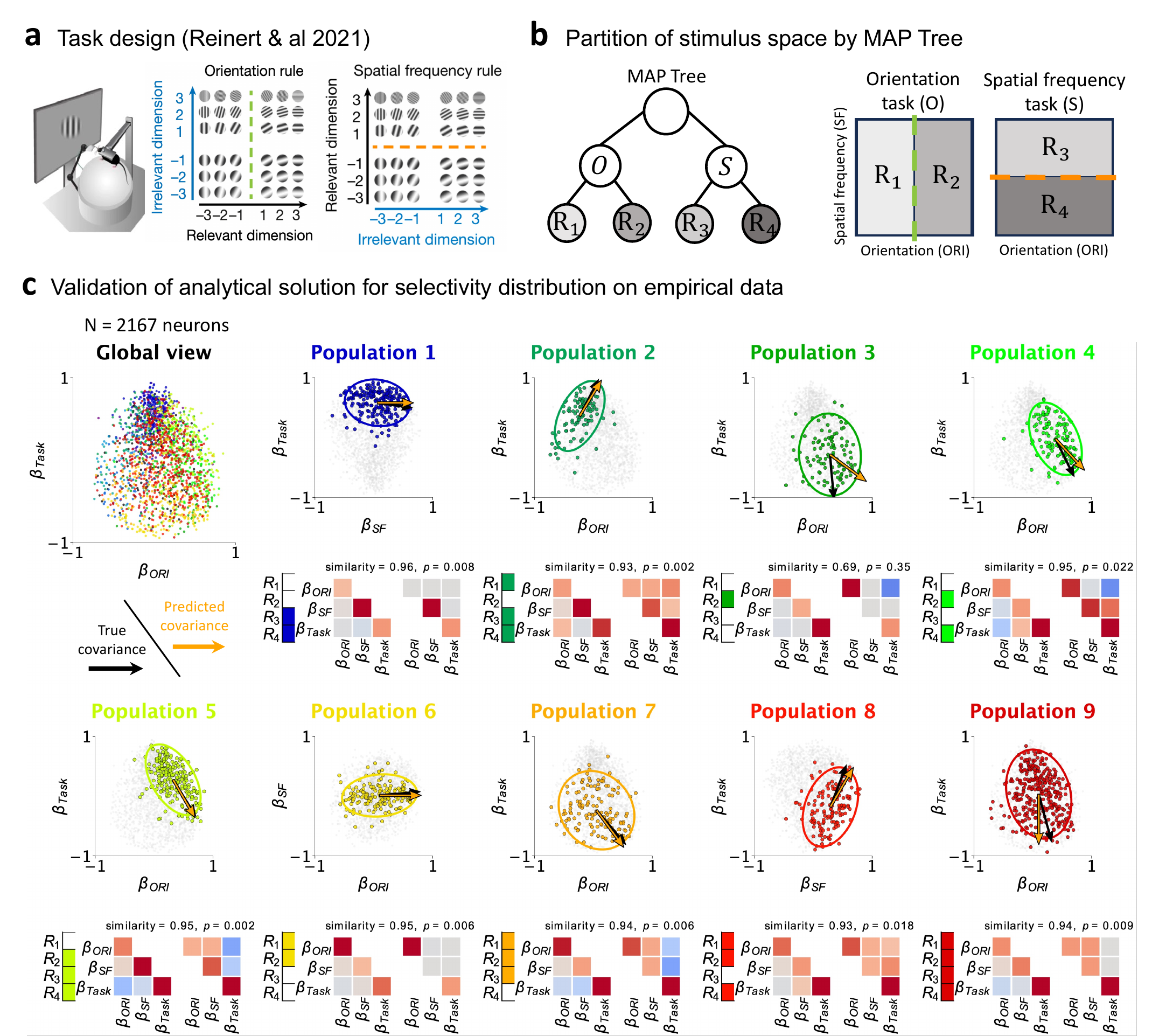}
\caption{\textbf{Task statistics predict the selectivity geometry of activation-defined prefrontal neural subpopulations.} 
\textbf{a,} Task design adapted from Reinert et al. (2021). Mice discriminated stimuli according to either
their orientation or their spatial frequency. 
Under the orientation rule, orientation was behaviorally relevant and spatial frequency was irrelevant; under
the spatial-frequency rule, these roles were reversed. Green and orange dashed lines indicate the category
boundary along the task-relevant stimulus dimension. 
\textbf{b,} Partition of the joint stimulus space by a MAP tree fitted to population activity. The first
split separates the orientation (O) and spatial-frequency (S) task contexts; subsequent splits along the
relevant stimulus dimension define four activation regions ($R_1$, $R_2$, $R_3$, and $R_4$), shown in the
corresponding stimulus spaces. 
\textbf{c,} Validation of the predicted selectivity geometry in 2,167 recorded neurons. The global view shows
neurons in selectivity space, and the nine detailed plots show activation-defined subpopulations
corresponding to distinct combinations of MAP-tree regions. 
In each plot, colored points highlight the selected subpopulation against the remaining neurons in gray, and
the ellipse represents its measured covariance in the displayed pair of selectivity dimensions. 
Black and orange arrows indicate the principal directions of the measured and theoretically predicted
covariance, respectively. The adjacent $R_1$-$R_4$ indicators mark in color the activation regions defining
each subpopulation. 
Covariance matrices compare the measured lower triangle with the theoretically predicted upper triangle
across orientation, spatial-frequency, and task selectivities. Matrix entries are rescaled coefficients
ranging from $-1$ (blue) to $+1$ (red). The displayed similarity is the cosine similarity between the unique
entries of the measured and predicted covariance matrices; the associated $p$ value is obtained from the
rotation-based permutation test described in Methods.}
\label{fig:selectivity_reinert}
\end{figure}

We tested this prediction with three simulated recurrent networks performing distinct context-dependent tasks and two empirical datasets. 
Across the simulated tasks, context determined either which of two stimulus features was
used for categorization (Supplementary Fig.~\ref{fig:supp_selectivity_rs}), the mixture of features defining
the category boundary (Supplementary Fig.~\ref{fig:supp_selectivity_rotation}), or whether the features were
combined additively or multiplicatively (Supplementary Fig.~\ref{fig:supp_selectivity_sign}). The empirical analyses used
prefrontal recordings from mice switching between orientation and spatial-frequency discrimination
\cite{reinert_mouse_2021} (Fig.~\ref{fig:selectivity_reinert}) and neural recordings from mice using context to
select between visual and auditory evidence \cite{hajnal_shifts_2024} (Supplementary
Fig.~\ref{fig:supp_selectivity_hajnal}). Across all five analyses, MAP trees fitted to predict population
activation patterns from the input recovered the task boundaries without access to target outputs
(Supplementary Figs.~\ref{fig:supp_selectivity_rs}c, \ref{fig:supp_selectivity_rotation}c,
\ref{fig:supp_selectivity_sign}c, and \ref{fig:supp_selectivity_hajnal}b; Fig.~\ref{fig:selectivity_reinert}b).
In the first empirical dataset, we analyzed 2,167 prefrontal neurons (Fig.~\ref{fig:selectivity_reinert}a). Neurons were
grouped according to the set of the regions in which they were active, producing nine subpopulations 
with distinct distributions in orientation, spatial-frequency and task-selectivity space.

For each subpopulation, we predicted selectivity geometry from the input-output statistics pooled over its
associated regions and compared this prediction with the covariance measured from neural activity
(Fig.~\ref{fig:selectivity_reinert}c; Supplementary Figs.~\ref{fig:supp_selectivity_rs}d,
\ref{fig:supp_selectivity_rotation}d, \ref{fig:supp_selectivity_sign}d, and
\ref{fig:supp_selectivity_hajnal}c). Selectivity geometry varied markedly across subpopulations, 
but this variation was accurately captured by the task-derived predictions, which recovered both 
the covariance structure and its principal directions. The mean similarity across subpopulations was significant in all five analyses (\(p<10^{-3}\) in each).
Together, these results establish activation-region structure as the basis of modular selectivity. 
Neurons active across different sets of regions form distinct subpopulations whose geometry reflects the input-output statistics processed within those regions.

\section{Neural baseline controls the balance between coarse- and fine-grained learning regimes}

Our tree interpretation directly links single-neuron operating points to network-level expressivity. 
We use neural baseline to denote the input-independent bias that sets unit preactivations relative to the ReLU threshold. 
By determining how readily units cross this threshold as inputs vary, neural baseline controls the emergence of new 
activation patterns and, in turn, the number of distinct affine maps the network can implement.

\begin{figure}[H]
\centering
\includegraphics[width=1\linewidth]{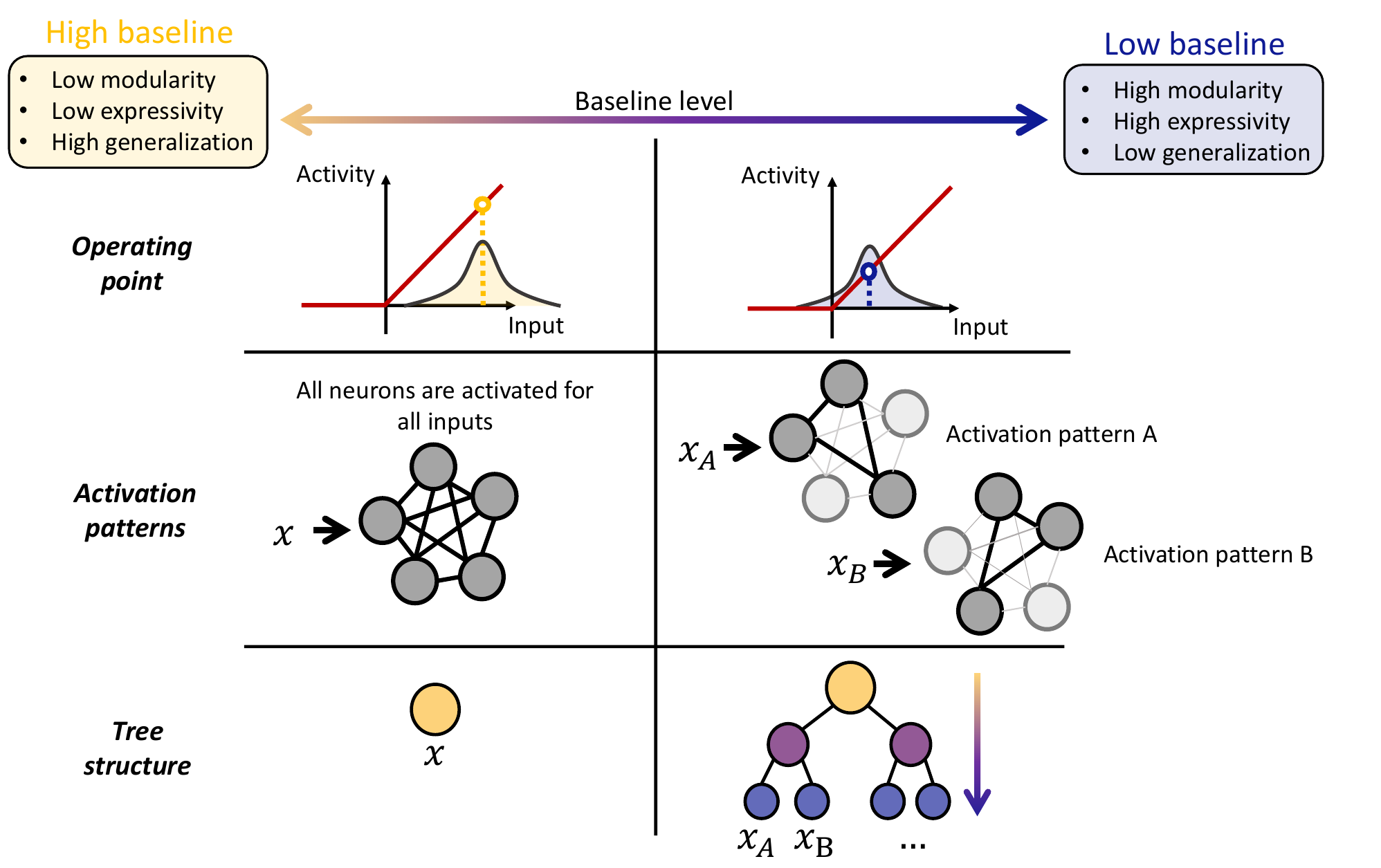}
\caption{\textbf{Neural baseline controls activation pattern diversity and the resolution of learned
input-output structure.} The horizontal continuum indicates a decrease from a high neural baseline (left,
yellow) to a low baseline closer to the activation threshold (right, blue). 
At a high baseline, the distribution of stimulus-driven inputs lies primarily on the active branch of the
rectified-linear activation function. The same units therefore remain active across inputs, producing a
common population activation pattern and routing the inputs to a single tree region. 
This regime implements few regional transformations and is predicted to have low modularity and expressivity
but stronger generalization. 
At a lower baseline, the stimulus distribution straddles the activation threshold, so changes in the input
switch subsets of units between active and inactive states. 
Different inputs, illustrated by $x_A$ and $x_B$, consequently recruit distinct population activation
patterns and are routed toward different leaves of a deeper tree. 
The downward arrow denotes the continued growth of the regional hierarchy as additional patterns emerge. 
This regime is predicted to support greater modularity and expressivity at the cost of reduced
generalization. 
Red curves show the common rectified-linear activation function; dashed vertical lines and colored circles
indicate the operating point in each regime; filled and unfilled neurons denote active and inactive units,
respectively.}
\label{fig:baseline_effect}
\end{figure}

When neural baselines lie far above threshold, most units remain active for all inputs
(Fig.~\ref{fig:baseline_effect}, left).
Their slope matrices are then nearly constant across the input space, so the network visits only a small
number of activation regions.
In the limiting case in which every unit remains on the same linear branch, $D_r = I$ and the regional
decomposition collapses to a single affine map. 
The corresponding MAP tree has one leaf, and the network implements a single regression plane. 
Despite having many parameters, its effective nonlinear expressivity is therefore low.

Moving the population baseline toward threshold has the opposite effect.
Small changes in the input can now switch individual units between their inactive and active states,
producing a larger collection of activation patterns across the input space (Fig.~\ref{fig:baseline_effect},
right). These additional patterns appear as new leaves in the effective tree implemented by the network, each
supporting a distinct local affine map.

This gain in expressivity for low-baseline networks may, however, come at a statistical cost.
Every additional leaf partitions the training set more finely, leaving fewer examples to constrain each
regional affine map. Finer regional partitions therefore support highly specialized solutions, but these
solutions are estimated from smaller effective sample sizes and are more sensitive to the particular examples
assigned to each region. New inputs may consequently be assigned to local models that received little nearby
training data. In this finite-data regime, increasing activation-pattern diversity can improve performance on
the training set while reducing the stability of predictions on unseen examples.

By contrast, a high-baseline network occupies fewer activation regions and pools more examples when estimating
each local map. Its shallower internal tree has lower expressivity but imposes greater sharing across the
dataset, favoring broad regularities over solutions specific to individual neighborhoods. This should reduce
estimation variance and improve generalization when data are limited or when the target function is
sufficiently smooth. The same constraint can nevertheless introduce approximation bias when the task requires
genuinely different computations in different parts of the input space. Neural baseline should therefore
control a bias-variance trade-off: moving toward threshold increases regional specialization, whereas moving
away from threshold favors simpler solutions supported by more observations. 
The best operating point should depend jointly on task nonlinearity and the amount of available training data.

We test these anticipated effects of neural baseline on network expressivity and generalization below.

\subsection{Activation-region formation enables nonlinear learning}

We first examined how neural baseline affects network expressivity during learning.
While a network remains confined to a single activation pattern, it implements one affine map and cannot
improve beyond the global least-squares solution. Reducing the loss below this limit requires a new activation
pattern and the additional regional transformation that it supports.

\begin{figure}[H]
\centering
\includegraphics[width=1\linewidth]{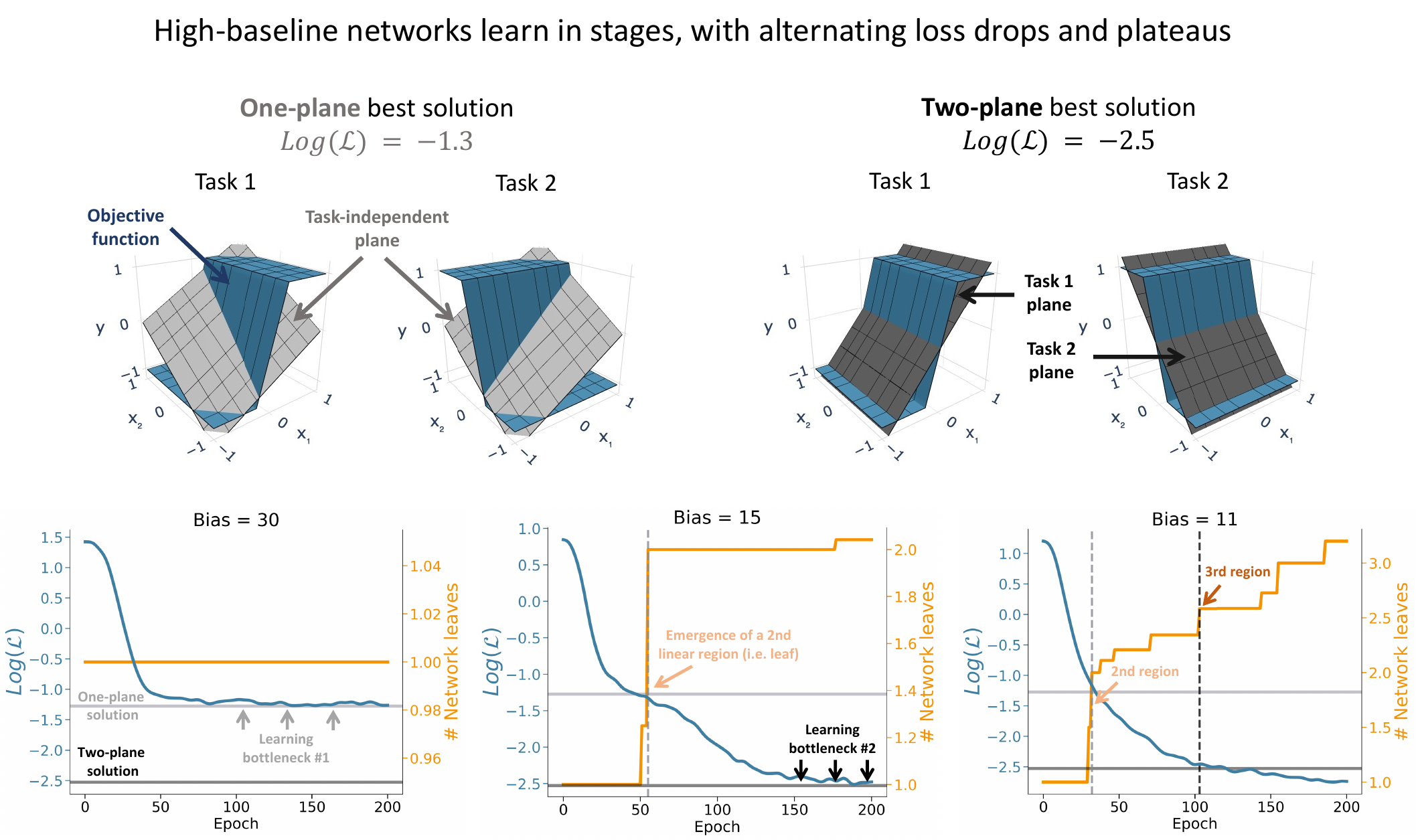}
\caption{\textbf{The emergence of activation regions enables learning beyond the global affine solution.}
\textbf{Top,} Geometric reference solutions for the rule-switch task. Blue surfaces show the task-dependent
target transformation and gray surfaces show the best affine approximation. A single task-independent plane
fitted jointly across both contexts yields the one-plane solution, with $\log(\mathcal L)=-1.3$. Fitting one
plane per task context yields the two-plane solution, with $\log(\mathcal L)=-2.5$. \textbf{Bottom,} Learning
trajectories for representative recurrent networks initialized with high ($b=30$), intermediate ($b=15$), or
lower ($b=11$) baselines. Blue curves show the logarithm of the training loss (left axes), and orange step
functions show the mean number of occupied MAP-tree leaves across epochs (right axes). Gray horizontal lines
indicate the errors of the one- and two-plane reference solutions. At $b=30$, the network remains within one
activation region and its loss plateaus near the one-plane solution. At $b=15$, the appearance of a second
activation region (dashed line) permits the loss to cross the one-plane boundary and approach the two-plane
solution. At $b=11$, successive appearances of second and third regions (dashed lines) are followed by
additional decreases in loss. Gray and black arrows mark learning plateaus or bottlenecks. Together, the
examples show that improvements beyond an affine reference solution coincide with increases in the number of
regional transformations available to the network.}
\label{fig:stage_learning}
\end{figure}

To test this prediction, we trained networks at different baselines and benchmarked their learning trajectories against two reference models: 
a one-plane solution fitted jointly across task contexts and a two-plane solution fitted separately within each context (Fig.~\ref{fig:stage_learning}, top). 
Learning unfolded in discrete stages that matched changes in activation-region structure (Fig.~\ref{fig:stage_learning}, bottom). 
At the highest baseline, the network occupied a single region and its loss plateaued near the one-plane solution. 
At an intermediate baseline, the emergence of a second occupied region coincided with the loss falling below the one-plane solution and approaching the two-plane solution.
Finally, at the lowest baseline, the formation of a third region coincided with the loss falling below the two-plane limit.
These trajectories show that the emergence of new activation regions enables successive reductions in loss beyond the limits of simpler affine solutions.

\subsection{Fine-grained learning comes at a generalization cost}

\begin{figure}[H]
\centering
\includegraphics[width=1\linewidth]{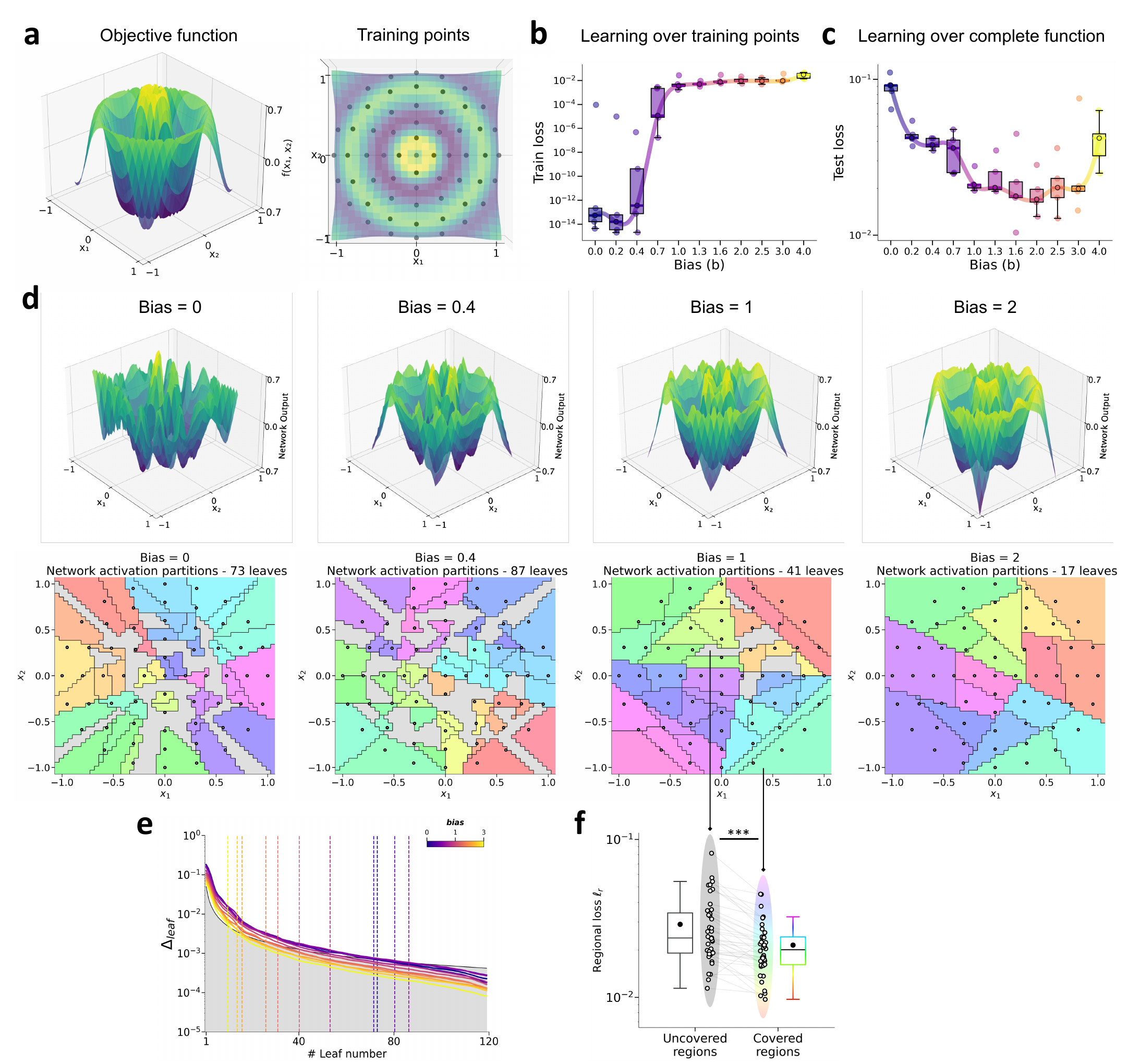}
\caption{\textbf{Neural baseline controls a trade-off between training fit and generalization.} \textbf{a,}
Nonlinear radial target function (left) and finite training set (right). Black points indicate the
training inputs distributed over six concentric circles; the colored surface shows the target value.
\textbf{b,} Final training loss as a function of neural baseline. The colored curve interpolates the median
across baselines. Colors progress from low (purple) to high (yellow) baseline. Low-baseline networks achieve
the smallest training losses. \textbf{c,} Final test loss on a dense held-out grid. In contrast to the
training loss, high-baseline networks generalize better to these new samples. \textbf{d,} Final network
output surfaces (top) and corresponding MAP-tree partitions of input space (bottom) for baselines $b=0$,
$0.4$, $1$, and $2$. Colors in the partition plots identify distinct leaves, black points show the training
inputs, and the number of occupied leaves is reported above each partition. Networks operating closer to
threshold form finer partitions and reproduce smaller-scale features of the target, whereas higher-baseline
networks form coarser partitions and smoother output surfaces. \textbf{e,} Marginal decrease in MAP-tree
cross-entropy, $\Delta_{\mathrm{leaf}}$, produced by each additional leaf. Curves show the mean across five
networks for each baseline; colored dashed lines indicate the estimated effective number of leaves, defined
by the point at which the incremental improvement falls below the depth-dependent threshold (black line; gray
area). \textbf{f,} Regional loss, $\ell_r$, within MAP-tree regions containing no training input (uncovered
regions) and regions containing at least one training input (covered regions). Mean losses are larger in uncovered
than covered regions (two-sided paired Wilcoxon signed-rank test, $n=55$ networks, $W=29$, $p=3.16\times10^{-9}$),
demonstrating the generalization cost of finely partitioning a finite training set. In \textbf{b},
\textbf{c}, and \textbf{f}, points denote individual network fits, boxes show the interquartile range and
median, and black circles show the mean.}
\label{fig:generalization}
\end{figure}

To assess how neural baseline affects generalization, we trained networks with different baselines to approximate the same nonlinear 
radial function from a sparse set of inputs distributed across concentric circles (Fig.~\ref{fig:generalization}a). 
Generalization was evaluated on a dense grid of previously unseen inputs spanning the full space. 
Networks operating closer to threshold fitted the training samples more accurately, whereas high-baseline 
networks achieved lower loss on the dense test grid (Fig.~\ref{fig:generalization}b,c).

The learned functions and their MAP-tree partitions clarified why baseline had opposite effects on training and test performance (Fig.~\ref{fig:generalization}d).
Low-baseline networks divided the input space into many regions and captured
small-scale variations of the target, whereas increasing baseline reduced the effective number of leaves and
produced smoother approximations. Consistent with this interpretation, MAP trees fitted to low-baseline networks 
required more leaves to capture activation patterns than those fitted to high-baseline networks (Fig.~\ref{fig:generalization}e). 
The finer partitions formed by low-baseline networks created regions containing no training samples, which we refer to as uncovered regions. 
Prediction error was significantly higher in uncovered than covered regions (\(W=29\), \(p=3.16\times10^{-9}\); Fig.~\ref{fig:generalization}f), 
indicating that poor generalization was concentrated in regions not visited by any training input.
Together, these results show that neural baseline sets the resolution of learning: lower baselines favor fine-grained representations, 
whereas higher baselines favor generalization beyond the training data.

\section{Discussion}

\paragraph{Inferring computation from representational geometry.}\leavevmode\\
A central contribution of our framework is to establish an analytical link between the function learned by a
network and the observable geometry of its internal representations. Starting from the network objective, we
derived necessary conditions on neural selectivity at learning stationarity. These constraints link each
neuron's selectivity to the input-output statistics of the regions in which it is active. Neurons active
across different sets of regions therefore develop distinct mixtures of feature selectivity, giving rise to
discrete subpopulations in selectivity space
\cite{rigotti_importance_2013,mante_context-dependent_2013,bernardi_geometry_2020,dubreuil_role_2022}.
Neural selectivity thus provides a measurable signature of the input-output relationships learned by the network.
Consistent with this prediction, the selectivity geometry derived from task statistics closely matched the 
geometry observed across neural subpopulations in both simulated and empirical datasets.
\\
\\
Our analysis therefore establishes a forward mapping from task structure to neural representation. More broadly, the
potential of this framework lies in \textit{reversing} this mapping: observable properties of neural activity 
could be used to infer the underlying computations performed by a network. 
Given the modular organization of a population and the selectivity geometry of its constituent subpopulations, one could seek to recover the
latent input-output relationships consistent with this geometry. This inverse problem will generally be
underdetermined, as distinct input-output relationships can produce equivalent selectivity covariances, for
example through rescaling or the influence of unobserved task variables. Nevertheless, the modular
organization of selectivity carries substantial information about the underlying computation and sharply
constrains the space of compatible functions. Extending the theory toward a full inverse formulation could
therefore turn representational geometry into a tool for functional identification, providing a principled
route from neural activity to the computations it implements.

\paragraph{The dual structure of nonlinear computation.}\leavevmode\\
At a more abstract level, our results place nonlinear neural networks at the interface between discrete and
continuous mathematics. Their activation structure partitions continuous input space into discrete regions,
while an affine map specifies the input-output transformation within each region. These two operations are
fundamentally different \cite{hahnloser_digital_2000,balestriero_spline_2018}. The discrete component selects the rule, whereas the continuous component implements
that rule. This distinction naturally separates input dimensions into modulators and drivers. Modulators,
such as task or contextual cues, move the network across regional boundaries and switch the transformation
being applied. Drivers, which may include stimulus features within a given region, continuously affect
neural activity without changing the selected transformation
\cite{salinas_transfer_1995,mante_context-dependent_2013,barbosa_early_2023}. In this sense, piecewise-linear networks derive
their expressivity from operating at the boundary between combinatorial and continuous systems.

\paragraph{Modularity in sensorimotor remapping tasks.}\leavevmode\\
We previously showed that sensorimotor remapping increases modularity in selectivity space \cite{tissot_sensorimotor_2026}. 
The present framework provides a mechanistic explanation for this observation. In these tasks, the same sensory
input must be mapped to different outputs depending on context, making the input-output relationship intrinsically nonlinear.
To implement this nonlinear mapping, the network must switch between distinct activation regions across contexts.
As a result, neurons recruited across different sets of activation regions become sensitive to different task statistics,
and form distinct subpopulations in selectivity space. Under this view, modular selectivity emerges directly from the regional 
decomposition required to implement a nonlinear mapping. The resulting modular organization should therefore depend on how 
readily distinct activation regions form during learning.

\paragraph{Neural baseline sets the resolution of learning.}\leavevmode\\
Our framework identifies neural baseline as a regulator of this process. By controlling activation-region diversity, 
baseline shapes the expressivity and modularity of learned representations and places networks along a 
continuum between two learning regimes.
At high baseline, most units remain active across inputs, limiting the diversity of activation patterns and
favoring a coarse representation of the input-output relationship. As baseline approaches the activation
threshold, inputs recruit increasingly distinct neural populations, enabling a more modular and fine-grained
representation. Our simulations support this transition from a ``big-picture'' learning regime, which captures broad
task structure, to a ``detail-oriented'' regime capable of dealing with finer variations.
\\
\\
An important next step will be to test this prediction empirically. The theory suggests that global inhibitory
or excitatory signals could shift biological networks between these learning regimes by modulating their operating
baseline \cite{chance_gain_2002}. Tasks requiring fine discrimination of the input-output relationship should benefit from moderate
inhibition that brings neural activity closer to threshold and increases the diversity of activation
patterns. By contrast, tasks dominated by coarse, large-scale structure may benefit from a higher-baseline
regime that limits unnecessary partitioning. These predictions directly link global modulation and representational modularity 
to the granularity of the input-output relationship that must be learned. Testing this prediction in biological 
networks will require a robust approach to identify activation regions from recorded activity.
\paragraph{Extending activation-region analysis to biological dynamics.}\leavevmode\\
In the analytical framework, activation regions have an exact definition: they comprise inputs processed 
by the same local affine transformation. To translate this decomposition into a practical analysis, we focused 
on ReLU networks, for which each unit has only two derivative states. A unit is active when its preactivation is positive, 
with a nonzero derivative, and inactive otherwise, with zero derivative. MAP trees recover this local derivative 
structure by predicting the binary activation state of every unit from the input, thereby providing an explicit
numerical approximation of the network's activation regions.
\\
\\
Applying the same construction to biological data requires an additional assumption because the activation
function of each recorded neuron is generally unknown. Neural activity must therefore
be discretized into putative active and inactive states, interpreted as regimes with nonzero and zero local
derivatives, respectively. This choice directly determines the inferred regional partition and its associated
neural subpopulations. Establishing biologically grounded discretization rules, adapted to the response
properties of the recorded cell types and circuits, will therefore be essential for strengthening the
application of the framework to empirical neural data.
\\
\\
Defining these states does not, however, ensure that regional membership is stable across trials.
Our derivation assumes deterministic, noise-free dynamics, whereby each input is assigned to a unique
activation pattern and a corresponding region of input space. In biological networks, trial-to-trial
variability breaks this one-to-one correspondence: repeated presentations of the same input can give rise to
different population activity patterns and, consequently, different regional assignments. Activation regions
should therefore be treated as stochastic rather than deterministic objects, characterized by input-dependent
probabilities over activation patterns \cite{arieli_dynamics_1996,churchland_stimulus_2010}. Incorporating this variability into the framework, for instance
through probabilistic region membership, will be an important step toward a more faithful description of
biological neural computation.
\\
\\
Our recurrent-network derivation further assumes that neural activity reaches equilibrium for each input.
This assumption may not hold in tasks with rapid temporal fluctuations, for which network activity
continuously evolves without approaching a fixed point
\cite{sussillo_opening_2013,chaisangmongkon_computing_2017}. The framework nevertheless extends naturally to
slowly varying inputs. When the characteristic timescale of input variation is long relative to the network's
relaxation time, activity remains close to the instantaneous equilibrium associated with the current input. A
time-dependent input-output relationship can then be treated as a sequence of quasi-stationary mappings,
allowing the regional decomposition to be applied locally in time. Extending the theory to regimes in which
input and network timescales are comparable will require an explicitly dynamical formulation. These extensions would describe 
variability around a regional partition, but leave open how that partition itself emerges and changes during learning.
\paragraph{Dynamics of activation regions during learning.}\leavevmode\\
Our derivations characterize the affine transformation associated with each activation region while treating
the regional partition as given. During learning, however, the transformations and the regions evolve
together: changes in network parameters move regional boundaries and can create, merge, or eliminate regions.
An analytical description of these dynamics would complete the present account of learning in nonlinear
networks. In particular, it would reveal how initial weights and neural baselines determine the trajectories
along which input space is partitioned, and why networks trained on the same task can converge to different
regional decompositions despite achieving similar performance \cite{driscoll_dynamic_2017,reinert_mouse_2021}.
\\
\\
Such a theory could also clarify variability in learning across individuals and training histories. Different
initial conditions may bias learners toward distinct partitions of the same task, producing alternative
strategies and representational geometries. The evolution of regions should also depend on how experience is
ordered. During sequential learning, partitions established by earlier tasks provide a pre-existing structure
that subsequent learning must reuse, refine, or reorganize, thereby shaping transfer and interference between
tasks \cite{yang_task_2019, kirkpatrick_overcoming_2017}. Under joint or continuously interleaved learning, by contrast, regional boundaries can co-evolve with
the statistics of all tasks. Describing these path-dependent regional dynamics would connect learning history
to the diversity and constraints of the representations that ultimately emerge.

\section{Methods}

\subsection{Network models and optimization}

Networks used the rectified-linear activation $\phi(z)=\max(0,z)$. The feedforward network comprised two
hidden layers and was defined by
\[
h^{(1)}=\phi\!\left(Vx+b_1\right),
\qquad
h^{(2)}=\phi\!\left(Wh^{(1)}+b_2\right),
\qquad
\widehat y=Uh^{(2)}.
\]
The recurrent networks received a constant input over the duration of each trial and followed the leaky
discrete-time dynamics
\[
h_t=\phi(\kappa_t),
\qquad
\kappa_{t+1}=(1-\alpha)\kappa_t+\alpha\left(Wh_t+Vx+b\right),
\qquad
\widehat y_t=Uh_t,
\]
where $\kappa_t$ is the recurrent state, $h_t$ the rectified activity, $W$ the recurrent matrix, $V$ the input
projection, $U$ the readout, and $b$ the additive recurrent bias. Because $b$ determines
the activity level of a unit in the absence of stimulus-driven fluctuations, we also refer to it as the
neural baseline. We used $\alpha=0.2$. Activity and output were summarized over the terminal window of length $\tau$ as
\[
\bar h=\frac{1}{\tau}\sum_{t=T-\tau+1}^{T}h_t,
\qquad
\bar y=\frac{1}{\tau}\sum_{t=T-\tau+1}^{T}\widehat y_t.
\]
We used 30 recurrent steps and $\tau=5$. Input and readout matrices were initialized
with orthogonal rows and held fixed throughout training, while the recurrent weight matrix was optimized. 
The initial weights were sampled as $W_{ij}\sim\mathcal N(0,\sigma_W^2)$, with $\sigma_W=\epsilon/\sqrt{m}$
for $m$ units with $\epsilon=10^{-4} << 1$ to ensure we are in a rich regime. The bias $b$ was optimized
during training unless explicitly held fixed, as in the experiments comparing prescribed baseline levels.

Networks were trained by minimizing
\[
\mathcal L_{\mathrm{MSE}}
=\frac{1}{p}\sum_{i=1}^{p}\left\lVert\widehat y_i-y_i\right\rVert_2^2
\]
with Adam \cite{kingma_adam_2017}, where $p$ is the number of training samples. We additionally included an activity regularization term to
capture the metabolic cost associated with neural activity \cite{attwell_energy_2001} and to place the network in the regime considered
in our derivation of neural selectivity. The training objective therefore was
\[
\mathcal L_{\mathrm{train}}
=
\mathcal L_{\mathrm{MSE}}
+
\lambda_h\frac{\lVert H\rVert_F^2}{N_H},
\]
where $H$ denotes the collection of hidden-unit activities over all training samples and, for recurrent
networks, over all recurrent time steps; $N_H$ is the total number of scalar entries in $H$. Gradients were clipped by their norm before
each optimizer step. 
The learning rate, activity penalty, and training duration are specified below for each experiment. 
No input or recurrent noise was added in the simulations reported here.

\subsection{Main Activation Pattern trees}

The Main Activation Pattern (MAP) tree is a multi-output oblique decision tree fitted post hoc to approximate
the routing performed by the neural network. Let $x_i\in\mathbb R^{n_x}$ denote the input on trial $i$ and
$a_i=(a_{i1},\ldots,a_{im})\in\{0,1\}^m$ its population activation pattern. The training data for the tree
are therefore
\[
X=(x_1^T,\ldots,x_p^T)^T\in\mathbb R^{p\times n_x},
\qquad
A=(a_1^T,\ldots,a_p^T)^T\in\{0,1\}^{p\times m}.
\]
Unlike an ordinary classification tree, which predicts a single class, the MAP tree jointly predicts the $m$
binary activation variables. Each internal node $v$ applies an oblique split
\[
x\in S_{v,L}\quad\Longleftrightarrow\quad w_v^Tx\leq t_v,
\qquad
x\in S_{v,R}\quad\Longleftrightarrow\quad w_v^Tx>t_v,
\]
where $S_v$ is the set of samples reaching the node, $w_v\in\mathbb R^{n_x}$ is a unit projection direction, and
$t_v$ is a scalar threshold. Thus, the leaves define a polyhedral partition of input space rather than a
partition restricted to axis-aligned rectangles.

For a sample set $S$, the empirical activation probability of unit $j$ is
\[
q_{S,j}=\frac{1}{|S|}\sum_{i\in S}a_{ij}.
\]
The node impurity used during fitting is the sample-weighted sum of the marginal binary entropies,
\[
I(S)=|S|\sum_{j=1}^{m}H_{\mathrm b}(q_{S,j}),
\qquad
H_{\mathrm b}(q)=-q\log q-(1-q)\log(1-q).
\]
For a candidate split $(w,t)$, the information gain is
\[
G(S;w,t)=I(S)-I(S_L)-I(S_R),
\]
and the pair $(w,t)$ producing the largest positive gain is retained. This marginal-entropy objective encourages each
leaf to contain a reproducible population activation pattern while avoiding the need to treat every one of
the potentially $2^m$ complete binary patterns as a separate class.

Candidate split directions were constructed at each node in three stages. First, we identified a candidate
direction in input space along which the population activation pattern varied most strongly. This direction
was estimated by ridge regression from the inputs to the binary activation patterns,
\[
B_v=(X_v^TX_v+\lambda I)^{-1}X_v^TA_v,
\qquad
\widetilde w_v=B_v\mathbf 1_m,
\qquad
w_v^{\mathrm{reg}}=\frac{\widetilde w_v}{\lVert\widetilde w_v\rVert_2}.
\]
The ridge parameter was $\lambda=10^{-6}$. This direction, along with a set of normalized random directions,
was included among the candidate split directions.
\[
w=\frac{u}{\lVert u\rVert_2},
\qquad u\sim\mathcal N(0,I_{n_x}).
\]
For each candidate direction $w$, the inputs were projected as $z_i=w^Tx_i$ and ordered along this
one-dimensional axis. A possible split was evaluated between each pair of adjacent samples with different
projected values, using their midpoint as the threshold. The selected split was the direction-threshold pair
with the largest information gain,
\[
(w^*,t^*)=\operatorname*{argmax}_{w,t}G(S_v;w,t).
\]
Finally, the selected direction was refined by testing normalized local perturbations of the form
\[
w'=\frac{w^*+\eta u}{\lVert w^*+\eta u\rVert_2},
\qquad u\sim\mathcal N(0,I_{n_x}),
\]
where $\eta=0.15$ controls the refinement scale.
\\
\\
Trees were grown best-first. Every current leaf proposed its best admissible split, and the leaf with the
greatest information gain was divided next. Consequently, a tree with $L+1$ leaves was obtained from the
$L$-leaf tree by adding the single split that most reduced total impurity. Growth stopped at the requested
maximum number of leaves or when no positive-gain admissible split remained.
\\
\\
A leaf $\ell$ stores the vector of empirical activation probabilities
\[
\widehat q_\ell=\frac{1}{|S_\ell|}\sum_{i\in S_\ell}a_i,
\qquad
\widehat q(x)=\widehat q_{\ell(x)},
\]
where $\ell(x)$ is the leaf reached by input $x$. Tree fit was measured by the mean binary cross-entropy
\[
\mathcal L_{\mathrm{MAP}}
=-\frac{1}{pm}\sum_{i=1}^{p}\sum_{j=1}^{m}
\left[a_{ij}\log_2\widehat q_j(x_i)
+(1-a_{ij})\log_2\bigl(1-\widehat q_j(x_i)\bigr)\right],
\]
with probabilities bounded numerically away from zero and one. We considered two rules for converting the
probability vector of leaf $\ell$ into a binary activation pattern. Under a threshold rule,
\[
\widehat a_{\ell,j}(\theta)
=\mathbb I\!\left[\widehat q_{\ell,j}>\theta\right],
\]
where $\theta$ is a fixed probability threshold. As a nonparametric alternative, the top-$k$
rule preserves the rounded expected number of active neurons in each leaf. Specifically, we set
\[
k_\ell=
\left\lfloor\sum_{j=1}^{m}\widehat q_{\ell,j}\right\rceil,
\qquad
\widehat a_{\ell,j}
=\mathbb I\!\left[j\in\operatorname{Top}_{k_\ell}(\widehat q_\ell)\right],
\]
where $\lfloor\cdot\rceil$ denotes the nearest integer and $\operatorname{Top}_{k}(q)$ the indices of the $k$
largest entries of $q$. 
Samples with the same resulting mask were treated as belonging to the same approximated activation region. 
The top-$k$ rule provides a threshold-free alternative when a fixed cutoff is undesirable.
Regional analyses excluded regions that did not contain enough samples or whose input matrix was numerically
ill conditioned, as detailed below.

The conversion of continuous activity into the binary targets $A$ differed between artificial ReLU networks
and recorded neural data. In artificial networks, activity has an explicit mechanistic threshold: unit $j$
was active precisely when its rectified activity was positive,
\[
a_{ij}=\mathbb I[h_j(x_i)>0].
\]
For feedforward networks, the activities of both hidden layers were concatenated before binarization, so that
the MAP-tree target contained the complete multilayer activation pattern. For recurrent networks, activity
was averaged over a terminal window of $\tau$ simulation steps,
\[
\bar h_j(x_i)=\frac{1}{\tau}\sum_{t=T-\tau+1}^{T}h_j(x_i,t),
\qquad
a_{ij}=\mathbb I[\bar h_j(x_i)>0],
\]
so that the tree described the stabilized end-of-trial computation rather than transient state changes. We
used $\tau=5$ throughout.

Unlike artificial ReLU networks, recorded neural populations do not provide a known activation function from
which binary activation states can be defined directly.
We therefore introduced a plausible binarization rule: each neuron’s spike rate was standardized across
trials, and responses above its mean firing rate were classified as active.
\[
z_{ij}=\frac{s_{ij}-\mu_j}{\sigma_j},
\qquad
\mu_j=\frac{1}{p}\sum_{i=1}^{p}s_{ij},
\qquad
a_{ij}=\mathbb I[z_{ij}>0].
\]
Thus, ``active'' denotes firing above that neuron's trial-averaged rate, rather than firing above an absolute
physiological threshold. This binary matrix was computed before the PCA-based processing used for the
separate continuous selectivity analysis. After this initial binarization, artificial and recorded data
followed the same MAP-tree fitting and binary-pattern prediction pipeline. Neurons with the same predicted
activation pattern across leaves were grouped into the activation-defined subpopulations used for the
selectivity analysis.

\subsection{Feedforward nonlinear regression and regional least squares}

The feedforward experiment in Fig.~\ref{fig:feedforward_simulation} used two hidden layers with 300 ReLU
units each, two inputs, and one linear output. The input and output projection matrices $V$ and $U$ were
initialized using orthogonal matrices and held fixed, while the hidden transformation $W$ and the initially
zero bias vectors $b_1$ and $b_2$ were trainable. The training set was a $40\times40$ regular grid over $[-1.5,1.5]^2$,
with target
\[
f(x_1,x_2)
=
\sin(2x_1)\cos(2x_2)
+0.3\exp\!\left[-\frac{x_1^2+x_2^2}{10}\right]
\sin(2x_1+2x_2).
\]
The network was trained for 1,000 optimizer steps using AdamW with learning rate $2\times10^{-3}$, and
activity regularizer $\lambda_h=10$. Parameters and MAP trees were saved every ten optimizer steps. At each
saved step, trees containing approximately logarithmically spaced numbers of leaves between 1 and 100 were
fitted with a minimum of ten samples per leaf.
\\
\\
For every MAP-tree region, the network slope $A_r$ and intercept $c_r$ were computed from the effective
weights associated with the predicted activation masks of the two hidden layers $D_r^{(1)}$ and $D_r^{(2)}$:
\[
A_r
=UD_r^{(2)}WD_r^{(1)}V,
\qquad
c_r
=UD_r^{(2)}\left(WD_r^{(1)}b_1+b_2\right).
\]
The corresponding affine map was written as $M_r=(c_r\mid A_r)$. Let $X_r$ and $Y_r$ contain the $p_r$
inputs and targets assigned to region $r$, and define
\[
\widetilde X_r=X_r-\bar x_r\mathbf 1_{p_r}^{T},
\qquad
\widetilde Y_r=Y_r-\bar y_r\mathbf 1_{p_r}^{T}.
\]
The regional affine least-squares slope and intercept were
\[
A_r^{LS}=\widetilde Y_r\widetilde X_r^T
\left(\widetilde X_r\widetilde X_r^T\right)^{-1},
\qquad
c_r^{LS}=\bar y_r-A_r^{LS}\bar x_r,
\]
and the corresponding affine map was $M_r^{LS}=(c_r^{LS}\mid A_r^{LS})$.
We quantified the deviation from the regional least-squares solution by $\Delta_r$, and the corresponding
mean prediction error by $\mathcal E_r$:
\[
\Delta_r=M_r-M_r^{LS},
\qquad
\mathcal E_r=\frac{1}{p_r}\sum_{i\in S_r}
\left\lVert\widehat y_i-y_i\right\rVert_2^2.
\]
A region was retained when it contained at least ten samples and its centered input matrix had full input rank.
To test the predicted square-root scaling, regional values from the largest fitted trees were pooled across training stages and fitted with
\[
\log_{10}\lVert\Delta_r\rVert_F
=\beta_0+\beta_1\log_{10}\mathcal E_r+\varepsilon_r.
\]
The slope, Pearson correlation coefficient, and two-sided $p$ value are reported in
Fig.~\ref{fig:feedforward_simulation}.

\subsection{Recurrent rule-switch task and fixed-point analysis}

The context $c\in\{-1,1\}$ selected the relevant stimulus coordinate. The decision target was
\[
y(x_1,x_2,c)=
\begin{cases}
\operatorname{sign}(x_1), & c=-1,\\
\operatorname{sign}(x_2), & c=1,
\end{cases}
\]
with zero assigned to the positive class.

The recurrent fixed-point experiment in Fig.~\ref{fig:rnn_simulation} used 40 ReLU units, three inputs
$(x_1,x_2,c)$, and two outputs; the first reported the decision target defined above and the second
reproduced the context. Stimulus coordinates were drawn uniformly from $[-1,1]$ and discretized into eight
bins, and each training batch contained 120 trials. Networks were trained for 4,000 optimizer steps using
AdamW with learning rate $2\times10^{-4}$ and activity regularizer $\lambda_h=1$. MAP trees with one to nine
leaves were fitted to activation patterns averaged over the terminal window of length $\tau$ defined above.

Fixed-point predictions were evaluated for a trained network with baseline $b=5$ using an eight-leaf MAP
tree. For input $x_i$ and predicted binary activation mask $a_i$, the region-specific equilibrium was
computed by solving the masked linear system
\[
\hat h_i
=
D_i\bigl(I-WD_i\bigr)^{-1}
\bigl(Vx_i+b\bigr),
\qquad D_i=\operatorname{diag}(a_i).
\]
The global comparison omitted the mask and used the single linear solution obtained by setting $D_i=I$.
Predictions were compared with the simulated activity at the final time step. Unit-wise accuracy was
summarized by
$R^2=1-\mathrm{MSE}/\operatorname{Var}(h)$, and prediction errors were summarized across units by the
natural logarithm of the mean squared error over trials. Principal-component displays were obtained by
fitting PCA to the complete set of simulated trajectories and projecting both observed and predicted
equilibria into the same space.

For the selectivity analysis, we used a separate rule-switch setup matching the sign and rotation experiments
described below. The recurrent network contained 1,000 ReLU units, three inputs $(x_1,x_2,c)$ and one linear
output. Inputs were arranged on a homogeneous $16\times16$ grid over $[-1,1]^2$,
repeated identically for both contexts. Each input was held constant throughout the recurrent sequence. The
network was trained for 200 optimizer steps using AdamW with learning rate $10^{-3}$ and activity regularizer
$\lambda_h=10$. The task loss was the mean squared error over the final five recurrent steps.

\subsection{Context-dependent sign task}

The sign-task experiment used a recurrent network with 1,000 ReLU units, three inputs $(x_1,x_2,c)$ and one
linear output. The context $c\in\{-1,1\}$ selected between two nonlinear
classification rules. The target was
\[
y(x_1,x_2,c)=
\begin{cases}
\operatorname{sign}(x_1+x_2), & c=-1,\\
\operatorname{sign}(x_1x_2), & c=1,
\end{cases}
\]
with zero assigned to the positive class. Inputs were arranged on a homogeneous $16\times16$ grid over
$[-1,1]^2$, repeated identically in both contexts, yielding 512 task conditions. Each input was held constant
throughout the recurrent sequence. The network was trained for 200 optimizer steps using AdamW with learning
rate $10^{-3}$ and activity regularizer $\lambda_h=10$. The task loss was the mean squared error over the
final five recurrent steps.

\subsection{Context-dependent rotation task}

The rotation-task experiment used the same recurrent architecture, input grid, and training procedure. Here,
context selected the orientation of a linear decision boundary. For context-dependent angles
$\theta_{-1}=0^\circ$ and $\theta_{+1}=60^\circ$, the target was
\[
y(x_1,x_2,c)
=\operatorname{sign}\!\left(
x_1\cos\theta_c+x_2\sin\theta_c
\right),
\qquad c\in\{-1,1\},
\]
again assigning zero to the positive class. Thus, the relevant input direction was rotated by $60^\circ$
between contexts while the sampled input distribution remained unchanged.

\subsection{Baseline-dependent expressivity}

To study how the operating point controls the number of activation regions, the rule-switch networks
described above were trained from five independent initializations at baselines spanning the range used in
the figures. For the systematic sweep, 40-unit networks were trained at 11 evenly spaced baselines between 0
and 5; additional high-baseline examples were trained to illustrate the staged learning regime. At each saved
training stage, MAP trees were fitted and their cross-entropy was measured as a function of leaf number $L$.
The improvement produced by adding one leaf was
\[
\Delta_{\mathrm{leaf}}(L)
=\mathcal L_{\mathrm{MAP}}(L)-\mathcal L_{\mathrm{MAP}}(L+1).
\]
The effective number of regions was defined by the first addition for which $\Delta_{\mathrm{leaf}}(L)<\delta$; the baseline sweep used $\delta=0.03$. 
Global one-plane and context-specific two-plane reference errors were obtained by ordinary
least squares, either across all task conditions or separately within the two contexts. These reference
errors identified plateaus at which a network behaved as one global affine transformation or as two
context-specific transformations.

\subsection{Finite-sample generalization experiment}

The generalization experiment in Fig.~\ref{fig:generalization} used recurrent networks with 300 ReLU units,
two inputs and one output. The nonlinear radial target was
\[
f(x_1,x_2)
=
\frac{\sin\!\left(10\sqrt{x_1^2+x_2^2}\right)}
{1+\sqrt{x_1^2+x_2^2}},
\]
The training set comprised 61 points arranged on six concentric circles with radii evenly spaced between 0
and 1 and with $1$, $4$, $8$, $12$, $16$, and $20$ equally spaced angular samples, respectively. Networks
were trained for 1,000 steps with AdamW, learning rate $10^{-3}$, no activity penalty. Eleven baseline
values, $b\in\{0,0.2,0.4,0.7,1,1.3,1.6,2,2.5,3,4\}$, were tested with five independent network
initializations per value.

Generalization error was evaluated on a $50\times50$ uniform grid $\mathcal G$ over $[-0.8,0.8]^2$ as
\[
\mathcal E_{\mathrm{gen}}
=\frac{1}{|\mathcal G|}\sum_{x\in\mathcal G}
\left\lVert\bar y(x)-f(x)\right\rVert_2^2,
\]
where $\bar y$ is the terminal-window average defined above. MAP trees with 1-120 leaves and at least five
samples per leaf were fitted. A leaf $\ell$ was designated trained when it contained at least one training
input,
\[
\chi_\ell=\mathbb I\!\left[S_\ell\cap\mathcal X_{\mathrm{train}}\neq\emptyset\right].
\]
Errors were averaged separately over leaves with $\chi_\ell=1$ and $\chi_\ell=0$. The paired comparison in
Fig.~\ref{fig:generalization} used a two-sided Wilcoxon signed-rank test across matched network fits ($W=29$,
$p=3.16\times10^{-9}$). Curves and shaded bands show the mean and mean $\pm$ one standard error across the
five initializations.

\subsection{Empirical data pre-processing}

\subsubsection{Reinert dataset}

Recordings from different animals and sessions were combined into a pseudo-population. Trials were divided
into eight conditions defined by task context, stimulus category, and behavioral outcome (hit, miss, correct
rejection, or false alarm). For each condition, the number of pseudo-trials was set to the mean number of
available trials across recordings. Trials were then sampled with replacement within each recording and
condition, and the resulting activity vectors were concatenated across recordings.

Let $H\in\mathbb R^{p\times n}$ denote the pseudo-trial-by-neuron spike-rate matrix. Each neuron's activity
was standardized across pseudo-trials as
\[
\widetilde H_{ij}=\frac{H_{ij}-\mu_{H,j}}{\sigma_{H,j}},
\]
where $\mu_{H,j}$ and $\sigma_{H,j}$ are the mean and standard deviation of neuron $j$ across pseudo-trials.
The binary activation-pattern matrix $A\in\{0,1\}^{p\times n}$ was then defined before PCA denoising by
\[
A_{ij}=\mathbb I[\widetilde H_{ij}>0],
\]
where $A_{ij}=1$ indicates that neuron $j$ was active on pseudo-trial $i$. For the continuous selectivity
analysis, $\widetilde H$ was transformed onto its first ten principal components and inverse transformed in
the original neuron space, thereby retaining the dominant shared population activity modes.

Let $o_i,s_i\in\{-1,+1\}$ denote the orientation and spatial-frequency categories on pseudo-trial $i$,
respectively, and let $c_i$ denote its task context. The input and target vectors were
\[
x_i=
\begin{bmatrix}
o_i\mathbb I[c_i=\mathrm{ORI}] &
s_i\mathbb I[c_i=\mathrm{SF}] &
c_i
\end{bmatrix}^{T},
\qquad
y_i=
\begin{bmatrix}
o_i\mathbb I[c_i=\mathrm{ORI}]
+s_i\mathbb I[c_i=\mathrm{SF}] & c_i
\end{bmatrix}^{T}.
\]
The first two input coordinates describe context-gated orientation and spatial-frequency evidence. The first
target coordinate encodes the context-dependent NoGo/Go action as $-1/+1$, and the second encodes context.
The columns of the resulting input and target matrices $X$ and $Y$ were standardized across pseudo-trials as
\[
\widetilde X_{ij}=\frac{X_{ij}-\mu_{X,j}}{\sigma_{X,j}},
\qquad
\widetilde Y_{ij}=\frac{Y_{ij}-\mu_{Y,j}}{\sigma_{Y,j}},
\]
where $\mu_{X,j}$ and $\sigma_{X,j}$, respectively $\mu_{Y,j}$ and $\sigma_{Y,j}$, are the mean and standard
deviation of column $j$ across pseudo-trials.

\subsubsection{Hajnal dataset}

Only correct trials from the complex task blocks were retained. Visual orientation ($45^\circ$ or
$135^\circ$), auditory frequency (5 or 10~kHz), and task context (auditory or visual) were encoded as binary
variables in $\{-1,+1\}$. ACC and V1 recordings were pooled for the analysis reported here. For each neuron
and trial, the response was defined as the mean firing rate during the 1.5-s interval following trial onset.

The three binary task variables defined eight experimental conditions. For each condition, the number of
pseudo-trials was set to the mean number of available trials across recordings. Each pseudo-trial was
constructed by independently sampling, with replacement, one response from the corresponding condition for
every neuron. The sampled responses were concatenated across neurons to form the pseudo-population activity
matrix $H$.

Each neuron's activity was standardized across pseudo-trials as
\[
\widetilde H_{ij}=\frac{H_{ij}-\mu_{H,j}}{\sigma_{H,j}},
\]
where $\mu_{H,j}$ and $\sigma_{H,j}$ are the mean and standard deviation of neuron $j$ across pseudo-trials.
The binary activation-pattern matrix $A\in\{0,1\}^{p\times n}$ was defined before PCA denoising by
\[
A_{ij}=\mathbb I[\widetilde H_{ij}>0].
\]
For the continuous selectivity analysis, $\widetilde H$ was transformed onto its first ten principal
components and inverse transformed in the original neuron space, thereby retaining the dominant shared
population activity modes.

Let $o_i,f_i\in\{-1,+1\}$ denote visual orientation and auditory frequency, respectively, and let $c_i$
denote task context. The input and target vectors were constructed in the same way as for the Reinert
dataset,
\[
x_i=
\begin{bmatrix}
o_i\mathbb I[c_i=\mathrm{visual}] &
f_i\mathbb I[c_i=\mathrm{auditory}] &
c_i
\end{bmatrix}^{T},
\qquad
y_i=
\begin{bmatrix}
o_i\mathbb I[c_i=\mathrm{visual}]
+f_i\mathbb I[c_i=\mathrm{auditory}] & c_i
\end{bmatrix}^{T}.
\]
Thus, the first target coordinate encoded the correct NoGo/Go action as $-1/+1$, and the second encoded
context. The columns of the resulting input and target matrices $X$ and $Y$ were standardized across
pseudo-trials as
\[
\widetilde X_{ij}=\frac{X_{ij}-\mu_{X,j}}{\sigma_{X,j}},
\qquad
\widetilde Y_{ij}=\frac{Y_{ij}-\mu_{Y,j}}{\sigma_{Y,j}},
\]
where $\mu_{X,j}$ and $\sigma_{X,j}$, respectively $\mu_{Y,j}$ and $\sigma_{Y,j}$, are the mean and standard
deviation of column $j$ across pseudo-trials.

\subsection{Neural-data selectivity analysis}

The selectivity coefficients were estimated from the standardized matrices by ordinary least squares,
\[
\widehat\beta=(\widetilde X^T\widetilde X)^{-1}
\widetilde X^T\widetilde H.
\]
Each column of \(\widehat{\beta}\) contains the selectivity coefficients of one neuron with respect to the
task variables represented in \(\widetilde X\).

A MAP tree was fitted to predict the binary activation-pattern matrix $A$ from the standardized task
variables $\widetilde X$, with at least ten pseudo-trials per leaf. The covariance analyses used four-leaf
trees throughout, except for the simulated sign task, for which a seven-leaf tree was used. Leaf
probabilities were converted into binary patterns using the threshold rule, with $\theta=0.1$ for the
three simulated tasks, $\theta=0.2$ for the Reinert dataset, and $\theta=0.3$ for the Hajnal dataset. For each
dataset, the threshold was selected to maximize the diversity of the resulting binary activation patterns.
Neurons were grouped into subpopulations according to the set of tree regions in which they were predicted
to be active.
Subpopulations with zero predicted covariance or containing less than 5\% of the recorded neurons were
excluded.

For a retained subpopulation $s$ containing $n_s$ neurons, let $\beta_j\in\mathbb R^3$ be the selectivity
vector of neuron $j$. Its measured selectivity geometry was
\[
\Sigma_s^{\mathrm{meas}}
=\frac{1}{n_s-1}\sum_{j\in s}
(\beta_j-\bar\beta_s)(\beta_j-\bar\beta_s)^T.
\]
To predict this geometry from task statistics, the standardized input variables $\widetilde X_s$ from leaves
in which the subpopulation was active were regressed onto the corresponding standardized task outputs
$\widetilde Y_s$,
\[
\beta_s=(\widetilde X_s^T\widetilde X_s)^{-1}
\widetilde X_s^T\widetilde Y_s,
\qquad
\Sigma_s^{\mathrm{pred}}=\beta_s\beta_s^T.
\]
Agreement was quantified by the cosine similarity between the vectors containing the unique entries of
$\Sigma_s^{\mathrm{meas}}$ and $\Sigma_s^{\mathrm{pred}}$. The corresponding similarity and rotation-test $p$
value are displayed for each subpopulation in Fig.~\ref{fig:selectivity_reinert}.

\subsection{Statistical reporting}

For the selectivity-geometry analysis, significance was assessed using Haar-distributed rotations. At each
permutation $k$, a $d\times d$ matrix $G_k$ was generated with independent entries
$(G_k)_{ij}\sim\mathcal N(0,1)$, where $d$ is the number of selectivity dimensions, and decomposed as
$G_k=Q_kR_k$ by QR
decomposition. The resulting matrix $Q_k$ satisfies $Q_k^TQ_k=I_d$ and defines a random rotation with no
preferred orientation \cite{mezzadri_how_2007}. The predicted covariance matrix of every retained subpopulation was then rotated as
$\Sigma_{s,k}^{\mathrm{null}}=Q_k\Sigma_s^{\mathrm{pred}}Q_k^T$. The same rotation was applied to all
subpopulations within a permutation.
If $z_s$ denotes the observed cosine similarity and $z_{s,k}^{\mathrm{null}}$ its value after rotation, the
population-dependent one-sided $p$ value was
\[
p_s=\frac{1+\sum_{k=1}^{K}\mathbb I[z_{s,k}^{\mathrm{null}}\geq z_s]}{K+1}.
\]
The aggregate statistic was the mean cosine similarity across the fixed set $\mathcal S$ of retained
subpopulations, $\bar z=|\mathcal S|^{-1}\sum_{s\in\mathcal S}z_s$. Its null distribution was obtained by
averaging across the same subpopulations after each common rotation, $\bar z_k^{\mathrm{null}}=|\mathcal
S|^{-1}\sum_{s\in\mathcal S}z_{s,k}^{\mathrm{null}}$, and its $p$ value was computed using the same
upper-tail formula. All analyses used $K=1{,}000$ rotations.

\clearpage
\section{Supplementary Figures}
\setcounter{figure}{0}
\renewcommand{\thefigure}{\arabic{figure}}
\renewcommand{\figurename}{Supplementary Fig.}

\begin{figure}[H]
\centering
\includegraphics[width=1\linewidth,trim=97bp 0 0 0,clip]{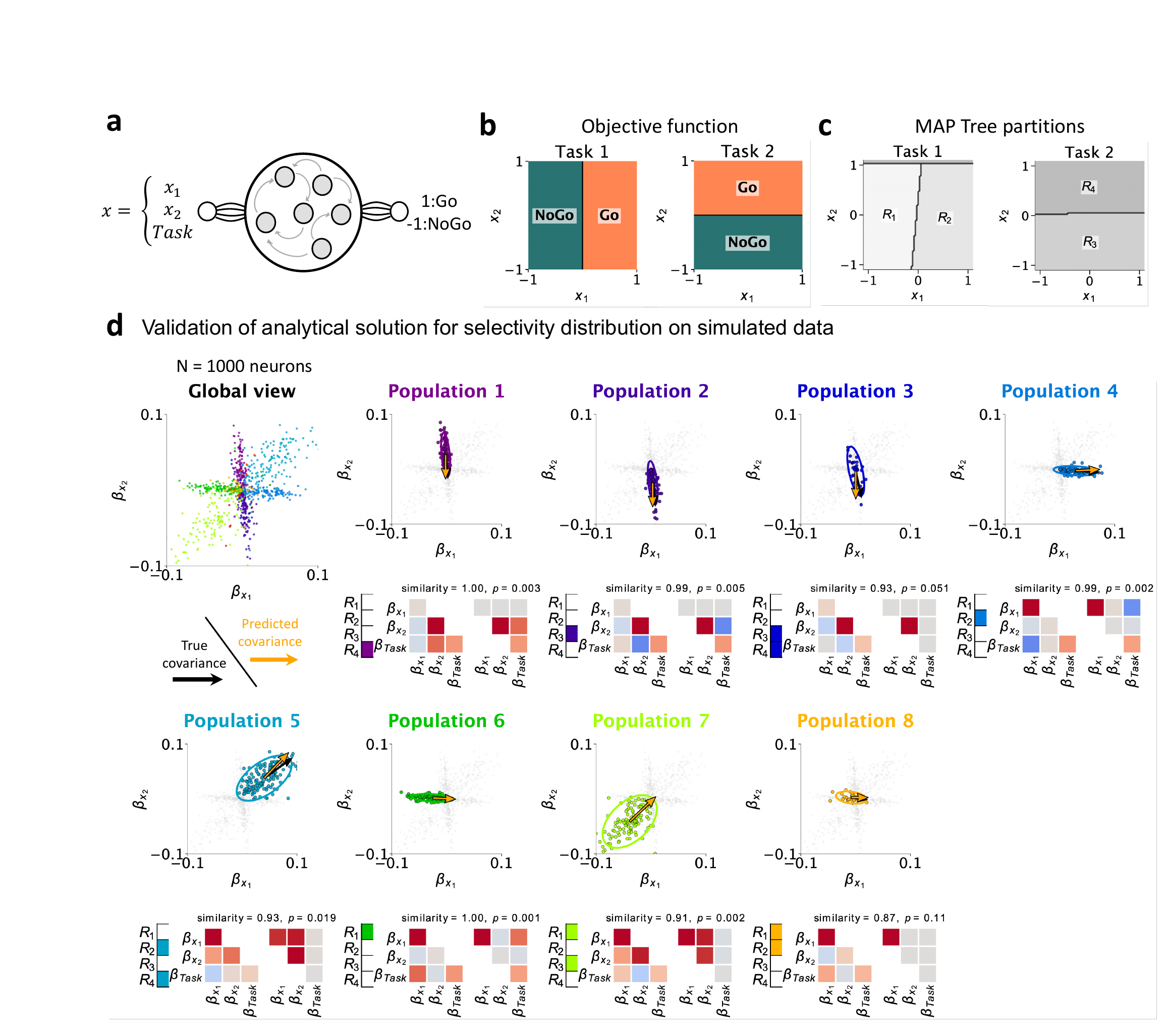}
\caption{\textbf{Task statistics predict the selectivity geometry of activation-defined neural subpopulations
in the simulated rule-switch task.}
\textbf{a,} Recurrent-network architecture. The network received two stimulus features, $x_1$ and $x_2$,
together with a task cue, and produced a binary go/no-go response.
\textbf{b,} Objective functions in the two task contexts. The response depended on the sign of $x_1$ in Task
1 and on the sign of $x_2$ in Task 2; black lines indicate the category boundaries.
\textbf{c,} Partition of the input space by a four-leaf MAP tree fitted to network activity. Gray levels
distinguish the regions $R_1$-$R_4$ in each task context.
\textbf{d,} Validation of the predicted selectivity geometry in 1,000 simulated neurons. The global view
shows all neurons in selectivity space, color-coded by activation-defined subpopulation, and the detailed
plots show each subpopulation separately. Colored points identify the selected subpopulation against the
remaining neurons in gray, and ellipses represent measured covariance. Black and orange arrows indicate the
principal directions of the measured and theoretically predicted covariance, respectively. The adjacent
region indicators identify the MAP-tree regions in which each subpopulation was active. Covariance matrices
compare the measured lower triangle with the theoretically predicted upper triangle across $x_1$, $x_2$, and
task selectivities. Matrix entries are rescaled coefficients ranging from $-1$ (blue) to $+1$ (red). The
displayed similarity is the cosine similarity between the unique entries of the measured and predicted
covariance matrices; the associated $p$ value is obtained from the rotation-based permutation test described
in Methods.}
\label{fig:supp_selectivity_rs}
\end{figure}

\begin{figure}[H]
\centering
\includegraphics[width=1\linewidth,trim=97bp 0 0 0,clip]{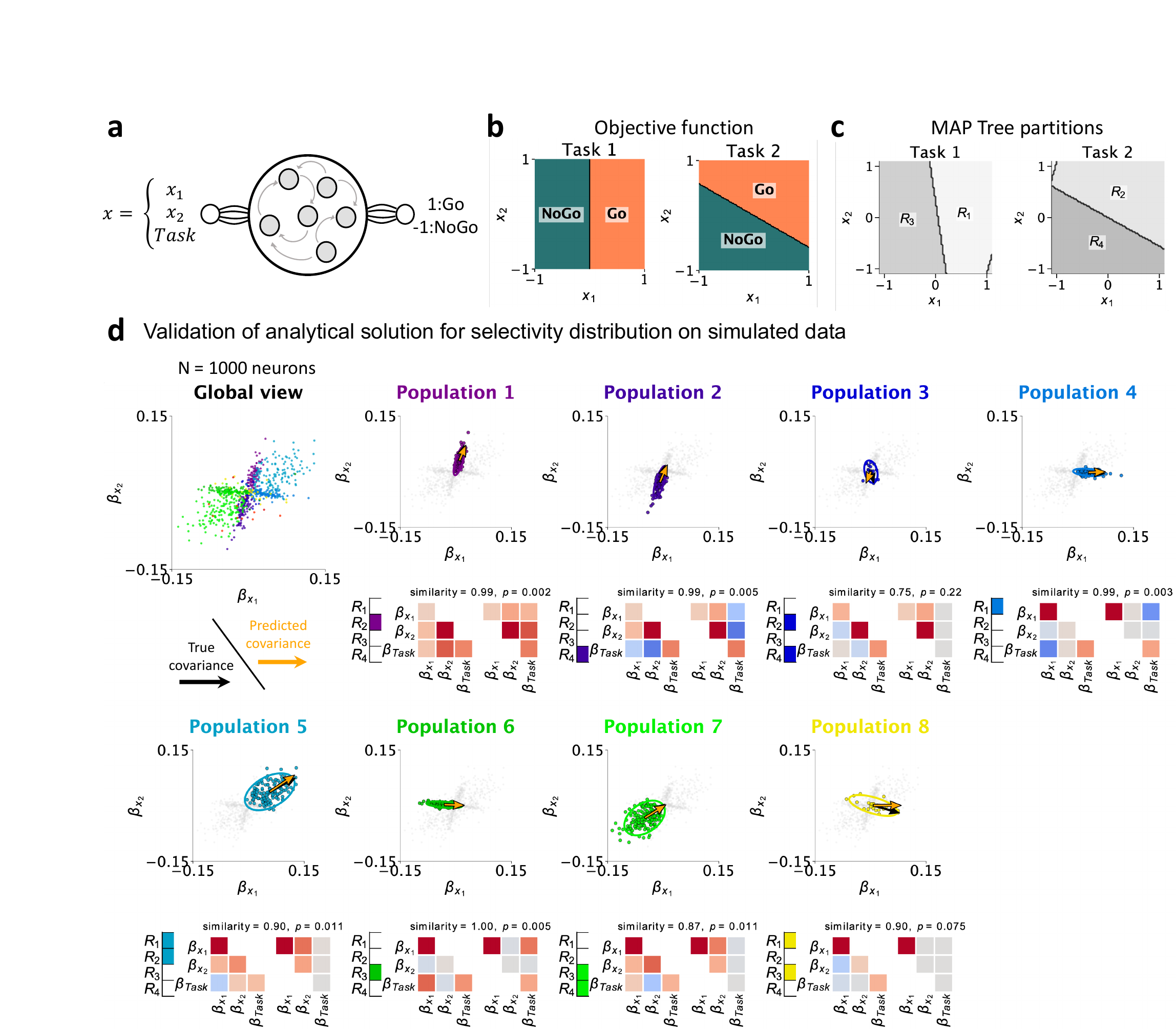}
\caption{\textbf{Task statistics predict the selectivity geometry of activation-defined neural subpopulations
in the simulated context-dependent rotation task.}
\textbf{a,} Recurrent-network architecture. The network received two stimulus features, $x_1$ and $x_2$,
together with a task cue, and produced a binary go/no-go response.
\textbf{b,} Objective functions in the two task contexts. The decision boundary was rotated from $0^\circ$ in
Task 1 to $60^\circ$ in Task 2; black lines indicate the category boundaries.
\textbf{c,} Partition of the input space by a four-leaf MAP tree fitted to network activity. Gray levels
distinguish the regions $R_1$-$R_4$ in each task context.
\textbf{d,} Validation of the predicted selectivity geometry in 1,000 simulated neurons. The global view
shows all neurons in selectivity space, color-coded by activation-defined subpopulation, and the detailed
plots show each subpopulation separately. Colored points identify the selected subpopulation against the
remaining neurons in gray, and ellipses represent measured covariance. Black and orange arrows indicate the
principal directions of the measured and theoretically predicted covariance, respectively. The adjacent
region indicators identify the MAP-tree regions in which each subpopulation was active. Covariance matrices
compare the measured lower triangle with the theoretically predicted upper triangle across $x_1$, $x_2$, and
task selectivities. Matrix entries are rescaled coefficients ranging from $-1$ (blue) to $+1$ (red). The
displayed similarity is the cosine similarity between the unique entries of the measured and predicted
covariance matrices; the associated $p$ value is obtained from the rotation-based permutation test described
in Methods.}
\label{fig:supp_selectivity_rotation}
\end{figure}

\begin{figure}[H]
\centering
\includegraphics[width=1\linewidth,trim=97bp 0 0 0,clip]{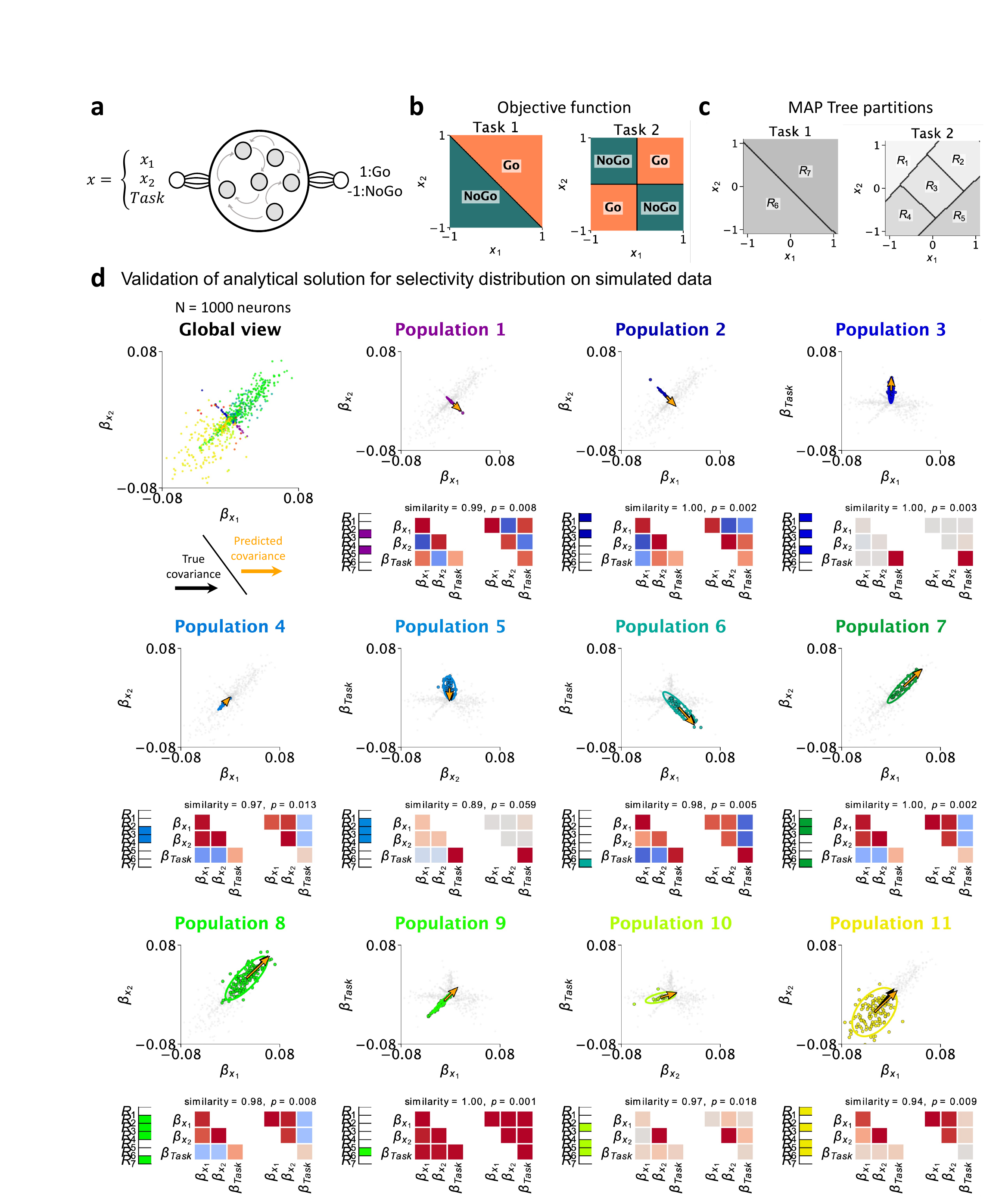}
\caption{\textbf{Task statistics predict the selectivity geometry of activation-defined neural subpopulations
in the simulated context-dependent sign task.} \textit{(Caption on next page.)}}
\label{fig:supp_selectivity_sign}
\end{figure}

\clearpage
\noindent\textbf{Supplementary Fig. 3. Task statistics predict the selectivity geometry of activation-defined
neural subpopulations in the simulated context-dependent sign task.}
\textbf{a,} Recurrent-network architecture. The network received two stimulus features, $x_1$ and $x_2$,
together with a task cue, and produced a binary go/no-go response.
\textbf{b,} Objective functions in the two task contexts. The response was determined by the sign of
$x_1+x_2$ in Task 1 and by the sign of $x_1x_2$ in Task 2; black lines indicate the category boundaries.
\textbf{c,} Partition of the input space by a seven-leaf MAP tree fitted to network activity. Gray levels
distinguish the regions $R_1$-$R_7$ in each task context.
\textbf{d,} Validation of the predicted selectivity geometry in 1,000 simulated neurons. The global view
shows all neurons in selectivity space, color-coded by activation-defined subpopulation, and the detailed
plots show each subpopulation separately. Colored points identify the selected subpopulation against the
remaining neurons in gray, and ellipses represent measured covariance. Black and orange arrows indicate the
principal directions of the measured and theoretically predicted covariance, respectively. The adjacent
region indicators identify the MAP-tree regions in which each subpopulation was active. Covariance matrices
compare the measured lower triangle with the theoretically predicted upper triangle across $x_1$, $x_2$, and
task selectivities. Matrix entries are rescaled coefficients ranging from $-1$ (blue) to $+1$ (red). The
displayed similarity is the cosine similarity between the unique entries of the measured and predicted
covariance matrices; the associated $p$ value is obtained from the rotation-based permutation test described
in Methods.

\begin{figure}[H]
\centering
\includegraphics[width=1\linewidth]{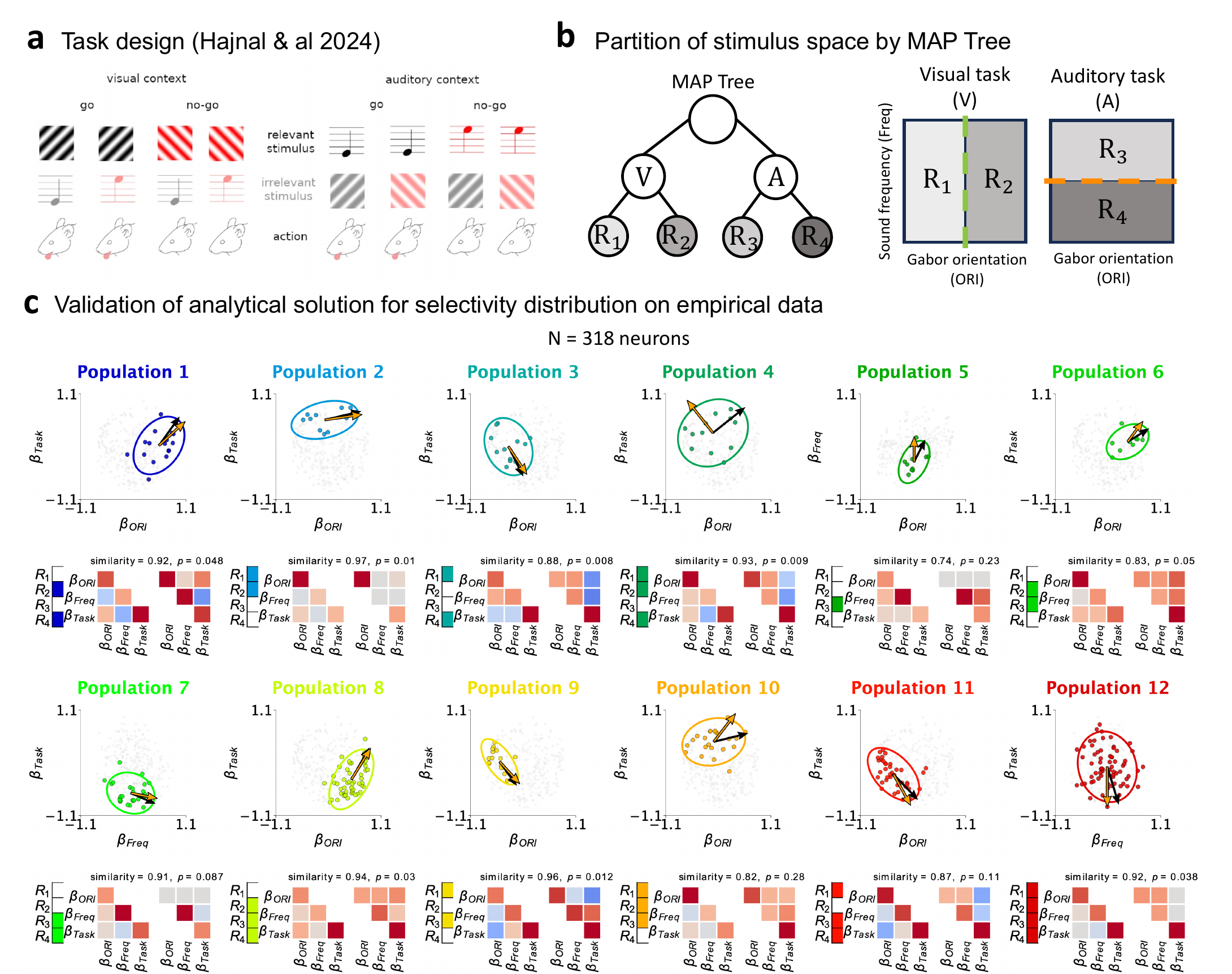}
\caption{\textbf{Task statistics predict the selectivity geometry of activation-defined neural subpopulations
in the Hajnal dataset.}
\textbf{a,} Task design adapted from Hajnal et al. (2024). Mice integrated visual orientation and auditory
frequency according to the current context to produce a go or no-go response.
\textbf{b,} Partition of the joint stimulus space by a MAP tree fitted to population activity. The first
split separates the visual (V) and auditory (A) contexts, and subsequent splits along the stimulus dimensions
define four activation regions ($R_1$-$R_4$), shown in the corresponding stimulus spaces. Green and orange
dashed lines indicate the category boundaries.
\textbf{c,} Validation of the predicted selectivity geometry in 318 recorded neurons. The twelve detailed
plots show activation-defined subpopulations corresponding to distinct combinations of MAP-tree regions. In
each plot, colored points highlight the selected subpopulation against the remaining neurons in gray, and the
ellipse represents its measured covariance in the displayed pair of selectivity dimensions. Black and orange
arrows indicate the principal directions of the measured and theoretically predicted covariance,
respectively. The adjacent $R_1$-$R_4$ indicators identify the activation regions defining each
subpopulation. Covariance matrices compare the measured lower triangle with the theoretically predicted upper
triangle across orientation, frequency, and task selectivities. Matrix entries are rescaled coefficients
ranging from $-1$ (blue) to $+1$ (red). The displayed similarity is the cosine similarity between the unique
entries of the measured and predicted covariance matrices; the associated $p$ value is obtained from the
rotation-based permutation test described in Methods.}
\label{fig:supp_selectivity_hajnal}
\end{figure}

\printbibliography

\end{document}